# An axon initial segment is required for temporal precision in action potential encoding by neuronal populations

Elinor Lazarov[1,2,3,4], Melanie Dannemeyer[3,5], Barbara Feulner[2,3], Jörg Enderlein[3,5], Michael J. Gutnick[1], Fred Wolf[2,3,6,7,8]*, Andreas Neef[2,3,6,7]*§

[1] Koret School of Veterinary Medicine, The Hebrew University of Jerusalem, P.O. Box 12, Rehovot 76100, Israel
[2] Max Planck Institute for Dynamics and Self-Organization, Am Faßberg 17, 37077 Göttingen, Germany
[3] Bernstein Center for Computational Neuroscience, Am Faßberg 17, 37077 Göttingen, Germany
[4] University Medical Center Göttingen, Department of Pediatrics and Adolescent Medicine, Division of Pediatric Neurology, Robert Koch Str.40, 37075 Göttingen, Germany
[5] III. Institute of Physics, Georg-August-University Göttingen, Friedrich Hund Pl. 1, 37077 Göttingen, Germany
[6] Institute for Nonlinear Dynamics, Georg-August University School of Science, Friedrich Hund Pl. 1, 37077 Göttingen, Germany
[7] Campus Institute for Dynamics of Biological Networks, Hermann Rein St. 3, 37075 Göttingen, Germany
[8] Max Planck Institute for Experimental Medicine, Hermann Rein St. 3, 37075 Göttingen, Germany
§ Lead Contact; * Corresponding authors: aneef@gwdg.de, fred@nld.ds.mpg.de

**Abstract**

Central neurons initiate action potentials (APs) in the axon initial segment (AIS), a compartment characterized by a high concentration of voltage-dependent ion channels and specialized cytoskeletal anchoring proteins arranged in a regular nanoscale pattern. Although the AIS was a key evolutionary innovation in neurons, the functional benefits it confers are not clear. Using a mutation of the AIS cytoskeletal protein βIV-spectrin, we here establish an *in vitro* model of neurons with a perturbed AIS architecture that retains nanoscale order but loses the ability to maintain a high $Na_V$ density. Combining experiments and simulations we show that a high $Na_V$ density in the AIS is not required for axonal AP initiation; it is however crucial for a high bandwidth of information encoding and AP timing precision. Our results provide the first experimental demonstration of axonal AP initiation without high axonal channel density and suggest that increasing the bandwidth of the neuronal code and hence the computational efficiency of network function was a major benefit of the evolution of the AIS.

## Introduction

Action potentials (AP) are the currency of information exchange between neurons in the brain. In central neurons, APs are initiated in the axon initial segment (AIS), a specialized compartment in the proximal axon. The AIS serves to separate the single output neurite - the axon - from the input, somato-dendritic surface of the cell (Hedstrom et al., 2008; Rasband, 2010). It extends over tens of micrometers and features a regular nanoscale arrangement of the cytoskeletal proteins, βIV-spectrin and AnkyrinG (AnkG) (Xu et al., 2013; Zhong et al., 2014). These are only found in the AIS and nodes of Ranvier (Bennett and Baines, 2001) and they control the creation and maintenance of the AIS (Hedstrom et al., 2008; Jenkins and Bennett, 2001; Komada and Soriano, 2002; Yang et al., 2007; Zhou et al., 1998). βIV-spectrin anchors AnkG to the cytoskeleton and AnkG, in turn, binds voltage-dependent sodium channels ($Na_V$) and potassium channels ($K_V$). These specializations emerged early in evolution. Although ancestral state reconstruction indicates that a giant Ankyrin isoform was already present in early bilaterians (Jegla et al., 2016), an AIS with characteristics similar to those of modern mammals first appeared in early chordates (Jenkins et al., 2015). First, $Na_V$ and later, $K_V$ channels acquired specific AnkG binding sites (Hill et al., 2008), thereby making the AIS a locus of ion channel clustering. The conservation of the AIS over more than 400 million years of vertebrate evolution suggests that it confers significant functional advantages on neurons and neural circuits. However, it is still not clear what these advantages are. The AIS establishes the proximal axon as the site of AP initiation by defining a single neuronal compartment where firing threshold is lowest. Since $Na_V$ density is highest in the AIS (Fleidervish et al., 2010; Kole et al., 2008; Lorincz and Nusser, 2010), it would be natural to conclude that it is the high channel density that makes the AIS the AP initiation site. However, it is important to note that additional factors contribute to the low axonal firing threshold, including, 1) a shifted voltage dependence of $Na_V$ in the AIS, which favors activation at less depolarized voltages than in the soma (Colbert and Pan, 2002; Hu et al., 2009; Kole et al., 2008), and 2) the large electrotonic distance of the AIS from the capacitive load of the soma, which renders distal $Na^+$-influx more efficient for *local* membrane depolarization (Baranauskas et al., 2013; Brette, 2013; Moore et al., 1983). Since APs originate in the AIS, the electrical properties of this compartment determine the precise timing of the APs in response to fluctuating synaptic inputs. Thus, the molecular architecture of the AIS directly influences a key process in neuronal information processing: the encoding of synaptic input, received and processed in the dendrites, into AP output. Indeed, theoretical studies predict that encoding will be strongly influenced by the density and properties of $Na_V$ channels (Fourcaud-Trocme et al., 2003; Naundorf et al., 2005).

In central neurons, $Na_V$ density in the AIS is larger than it is in the soma (Fleidervish et al., 2010; Kole et al., 2008). However, this does not necessarily mean that higher AIS channel density is prerequisite for axonal AP initiation; experimental evidence for this assertion is lacking. Computational simulations have been inconclusive. Some have indicated that axonal AP initiation is possible without any axon/soma gradient of channel density or channel properties (Baranauskas et al., 2013; Moore et al., 1983), while others have posited the need for an axonal channel density that is dozens of times higher than the somatic channel density (Kole et al., 2008; Mainen et al., 1995). Here, we experimentally address two questions crucial for understanding the benefits associated with AIS molecular architecture: 1) Is a high AIS channel density required for axonal AP initiation? and 2) Is a high AIS channel density required for precisely timed AP generation?

In order to answer these questions experimentally, it is necessary to manipulate the ratio of axon/soma ion channel density. However, in so doing, it is important to maintain a proper ratio of $Na_V$ and $K_V$ channels, since it is this ratio that determines the excitability of the AIS (Xu and Cooper, 2015; Zhou et al., 1998). Therefore, a pharmacological approach using specific channel blockers would be a problematic experimental strategy because the balance can hardly be reliably controlled.

Here, instead of pharmacology, we exploit a fundamental feature of AIS evolution: $Na_V$ and $K_V$ channels compete for a common binding region on the AnkG molecule (Xu and Cooper, 2015). As a consequence, the ratio between $Na_V$ and $K_V$ channels is determined by their relative affinity to AnkG, while the overall channel density scales with the number of AnkG molecules. We reduced the number of AIS ion channels indirectly through a mutation of the AnkG anchor, βIV-spectrin. The $qv^{3J}$ mutation results in a scrambled C-terminus due to a frame-shifting base insertion in the Spntb4 gene (Fig. 1C). It is the most conservative of all qv mutations and causes a mild, progressive phenotype that reflects a gradual change in properties of AIS and nodes of Ranvier over the course of postnatal development (Parkinson et al., 2001; Yang et al., 2007, 2004).

We now report that as $qv^{3J}$ neurons mature, loss of βIV-spectrin results in loss of AnkG and a marked decrease of axonal $Na_V$ density, while somatic $Na_V$ density is not different from control. Despite this shift in $Na_V$ density ratio, APs could still originate in the AIS. However, those APs were far less precisely timed and the bandwidth of AP encoding was substantially reduced. We conclude that improved encoding of information by precisely locking APs to dynamic inputs constitutes a major consequence and benefit of high $Na_V$ channel density in the AIS.

## Results

### Modeling predicts that APs can initiate in the axon even with low axonal $Na_V$ density

We first examined the predicted impact of reducing channel density at the AIS in multi compartment models of pyramidal neurons. An earlier study of such a model had found that APs initiate in the first node of Ranvier if the $Na_V$ density in the AIS is lowered (Kole et al., 2008). Hallermann, Kock, Stuart and Kole (2012) later devised an improved model of AP initiation, featuring updated $K_V$ and $Na_V$ channel models. Here, we used this latter model to study AP initiation as we reduce the ion channel density in the axon. We expect that removal of AnkG leads to a reduction of $Na_V$ and $K_V$ throughout the axon, but not the soma, where AnkG is not found. To mimic this, we used the same factor to reduce $Na_V$ and $K_V$ conductances in the axon and studied AP initiation in response to somatic current injections.

Even when the density of channels in the axon was lowered to the somatic level (Na conductance of 500 pS/µm²), APs still initiated in the AIS(Fig. S1B, C). This is evident from a comparison of the voltages at the soma and axon during AP onset. In addition, the non-somatic initiation can also be deduced from the waveform of the AP in the soma. During the earliest phase of an AP, when the AIS is more depolarized than the soma, the somato-dendritic membrane is charged by current flow from the AIS. This lateral current underlies the first phase of depolarization in the somatic AP waveform, and can be clearly seen in the phaseplot as an abruptly rising, early transient (Fig. 1 F, S1 A, 2 B). Unlike the axonal voltage, this feature of the somatic AP waveform can be readily measured in electrophysiological experiments and can be used to identify the axonal origin of APs. When we decreased the channel density in the AIS, this first phase of AP depolarization changed, but the second phase was only minimally affected (Fig S1 A). Furthermore, analysis of the somatic AP shape (see methods) showed that AP threshold became more depolarized and the initial slope of the phaseplot, the onset rapidness, was reduced (Fig. S1A). A similar pattern was observed for another detailed neuron model (Hay 2013, not shown).

The results of our simulations are in line with an early theoretical study of the AP initiation site in a less complex model (Moore et al., 1983) that demonstrated AP initiation in the AIS, if soma and axon had the same density of ion channels and the dendrites were passive. If the axon can remain the locus of AP initiation even when ion channel densities are lower than normal, what then is the functional importance of the high density that is maintained by the specialized AIS proteins?

**Modeling predicts that high axonal $Na_V$ density improves the temporal precision of spike encoding**

Interestingly, the family of qv mutations has been associated with an impaired AP timing precision. For example, various qv mutations are characterized by disruption of auditory evoked brainstem potentials, which reflect the synchrony of AP generation across a neuronal population (Parkinson et al., 2001). To test timing precision in the model, we measured the model's dynamic gain (Boucsein et al., 2009; Carandini et al., 1996; Higgs and Spain, 2009; Knight, 1972; Silberberg et al., 2004; Tchumatchenko et al., 2011; Touzel and Wolf, 2015). This function reveals how well the neuron can lock its APs to current fluctuations at different input frequencies in the presence of synaptic background activity. It expresses the susceptibility of the neuron's firing rate in Hz/nA across a range of input frequencies. The higher the bandwidth of the dynamic gain, the more precisely APs are locked to brief fluctuations in the input current.

We next examined the relationship between encoding bandwidth and AIS channel density using a model that most closely matches AIS ion channel kinetics (Hallermann et al., 2012) and in which the AIS locus of AP initiation was maintained even in the absence of a soma/AIS gradient in channel density. By contrast with the few seconds of simulated time needed for AP shape analysis (Fig. S1), calculation of a dynamic gain requires much more extensive simulation, as we collected $10^6$ APs for each condition in order to calculate the gain curves with high precision. For a morphologically realistic model, this would require years of processor time. Therefore, we sped up the computation by compacting the model's morphology (see Methods). This modification did not alter the AP initiation properties (compare Figs. S1 and 1). In the simulations, APs initiated in the AIS even when the channel densities in the axon were reduced by 95%, reversing the axon:soma gradient. Only a 99% reduction led to coincident initiation in AIS and soma, as evidenced in plots of axonal vs somatic voltage (Fig. 1E), and in the monophasic AP phaseplot (Fig. 1F, yellow trace). As the axonal channel density was reduced, the shape of the early somatic action potential waveform changed drastically. The threshold voltage shifted 20 mV to less negative values and the onset rapidness dropped from 29.8 to 4.5 $ms^{-1}$ (Fig. 1G).

Having assured, that the compacted, computationally less demanding model behaves very similar to the original model (Hallermann et al., 2012) we next used the compacted model to test, whether the bandwidth of information encoding is affected by a reduced axonal channel density, even if APs are still initiated in the AIS, i.e. with axonal channel densities corresponding to 0 to 90% reduction. We stimulated the model with fluctuating currents (Methods) similar to published electrophysiological experiments that measured the dynamic gain (Higgs and Spain, 2009). Mean and standard deviation of the stimulus were adjusted to obtain a firing rate of 2 Hz and a coefficient of variation of the inter-spike intervals of $0.95 \pm 0.03$, i.e. irregular firing. The correlation time of the stimulus was 35 ms. Following the established analysis procedure (Higgs and Spain, 2009) we obtained dynamic gain curves for models with different axonal channel densities (Methods, Fig. 1H). To unambiguously define the bandwidth of the gain curves, we identified the cut-off frequency as the frequency at which the gain dropped to 60% of its maximal value. This cut-off frequency became smaller, meaning the precision of AP timing deteriorated, as the axonal channel densities were reduced (Fig. 1I).

Our analysis suggests that the high axonal channel densities, created by a special molecular composition, are not important to achieve an axonal AP initiation locus, but instead are maintained to assure precise AP timing. Next, we set out to test whether these simulation results could be replicated in patch-clamp experiments.

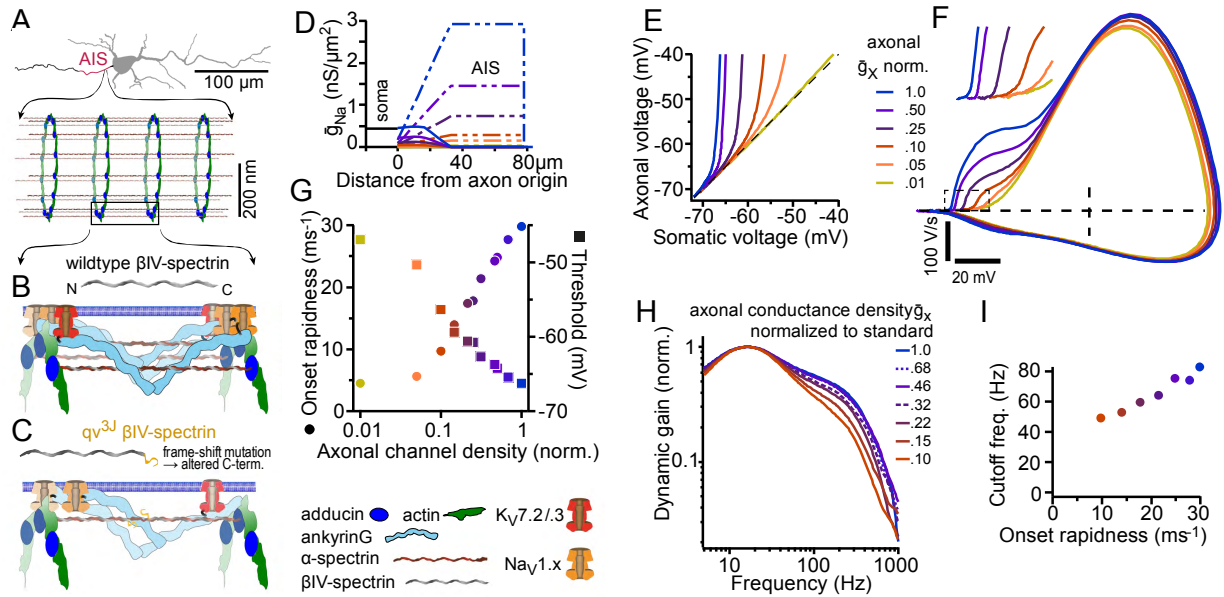

**Figure 1: Molecular down-regulation of AIS channel densities predicts axonal AP initiation with reduced timing precision** (A) The axon cytoskeleton in the AIS is highly regular. Tetramers of α- and β-spectrin serve as 190 nm long spacers between rings of actin (green) and adducin (blue) (Huang et al., 2017; Xu et al., 2013). (B) $Na_V$- and $K_V$-channels are anchored to AnkG (light blue), which binds to βIV-spectrin and the lipid membrane (blue double layer). (C) The qv$^{3J}$ mutation affects the very C-terminal portion of βIV-spectrin. We hypothesized that this might lead to a reduction in the number of intact tetramers and hence a reduction of AnkG and ion channels. The relative sizes of cytoskeletal proteins follows measures from electron microscopy (Jones et al., 2014). (D) We used a compacted version of a detailed biophysical model of pyramidal neurons (Hallermann et al., 2012) to study AP initiation under reduced AIS channel densities. The conductance densities of two Na channel subtypes in the model are displayed here (continuous lines – somatic, dashed lines – axonal channel variant with more hyperpolarized activation curve). All other conductances were also scaled down proportionally but are omitted here for clarity. Color code in D – I is identical, steps are identical in E and F. Note that already for a 90% reduction (red) the somatic conductance density is higher than the axonal one, for 95% reduction (orange) the soma:axon density ratio is 3:1. (E) The voltages occurring during AP onset in the soma and 50 µm into the axon are plotted against each other. Even when axonal channel density is lower than somatic, the AP still started in the axon. Only for a 99% reduction of channel densities (soma:axon density ratio 15:1) no sign of axonal initiation could be detected. (F) Phaseplots, plotting the first temporal derivative of the somatic membrane voltage $dV_m/dt$ against $V_m$, also showed a gradual change in the properties of the AP onset, as the axonal channel densities were reduced. Threshold was shifted by 20 mV and the initial lateral current into the soma was less pronounced; in particular the initial slope in the phase plot, called onset rapidness, decreased from 29.8 ms$^{-1}$ to 4.5 ms$^{-1}$. In contrast, the second phase of the AP waveform remained largely unchanged. Biphasic phase plots, indicating axonal initiation of APs, were obtained for density reduction as severe as 95%. (G) AP threshold (squares) and onset rapidness (circles) plotted against the degree of axonal channel density reduction. (H) Dynamic gain curves were calculated from 10$^6$ APs for each condition (see Methods). Dynamic gain curves showed a reduced bandwidth, when the channel densities in the axonal compartment were

reduced to 10% in six exponentially spaced steps. (I) Cut-off frequencies, defined as the frequencies, at which the gain reaches 60% of maximum, drop as the axonal channel density is reduced, even though for all densities probed, the AP starts in the soma, not the axon.

**The initial but not the somatic phase of APs is affected by qv$^{3J}$**

In order to test whether the simulation results were indeed predictive of actual neuron function, we studied the neurons from mice homozygous for qv$^{3J}$ (mutants) and control littermates throughout development, starting in the second week, just after the onset of AP firing, until the sixth week *in vitro*, using whole-cell patch clamp. During the first two weeks of this period, the membrane capacitance and conductance, assessed by short subthreshold current pulses, increased by about 50% and 100% respectively (Fig. 2A). This reflects a growth of the somato-dendritic compartment and occurred in both genotypes. We elicited APs by 100 ms current pulses, starting from a membrane voltage of -76 mV. The first AP evoked within a few milliseconds after current onset is analyzed as shown in the phase-plane plot in Fig. 2B and explained in the methods. The early phase, associated with AP initiation, is characterized by the threshold voltage and the onset rapidness (Fig. 2C, D). The second part of the AP upstroke is characterized by the peak rate of voltage rise, and the peak voltage (Fig. 2E and F).

We observed no effect of the qv$^{3J}$ mutation on passive cellular properties (Fig. 2A) and also no effect on those features of the AP waveform that are dominated by ion channels in the soma (Fig. 2E and F). The 95% confidence intervals of those properties in mutants and control overlap at almost all developmental stages. In contrast, the mutation showed a clear effect for AP properties that are dominated by axonal ion channels. In qv$^{3J}$ neurons AP initiation requires a more depolarized somatic potential (Fig. 2C) and it progresses more slowly (Fig. 2D). These observations mirror the outcome of the simulations for reduced axonal channel densities (Fig. 1F, G). The differences between the genotypes were small in the second week in cultures but during the next 4 weeks the they increased until the onset rapidness in mutants was only 48% of control (17±1 vs 8±1 ms$^{-1}$) and the threshold potential differed by 8 mV (-50 ±1 vs 42±1). The functional consequences of the $qv^{3J}$ mutation seem to be limited to early AP stages, which occur in the AIS, while the somatic phase of depolarization seems unaffected. These observations and the comparison to the simulations provide strong evidence for an unchanged development of somato-dendritic compartments. However, the nature of whole-cell patch clamp measurement does not directly provide the spatial resolution needed for more quantitative statements on the differences of axonal currents. To further investigate the effects of the $qv^{3J}$ mutation with spatial resolution, we turned to immunocytochemistry.

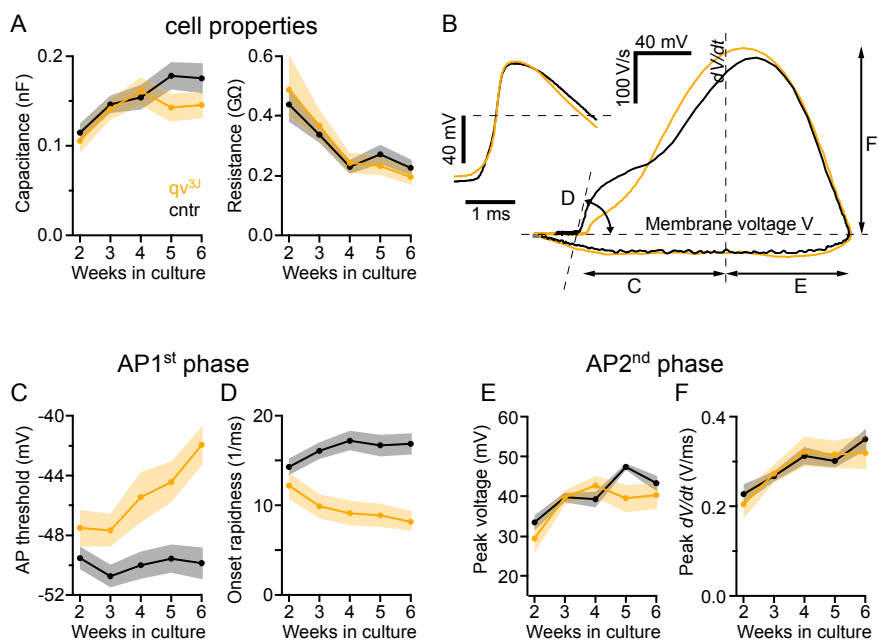

**Figure 2: APs of $qv^{3J}$ mutants show depolarized threshold voltage and slower onset rapidness compared to control.** (A) The $qv^{3J}$ mutation has no effect on the cell membrane properties. Cell capacitance and resistance are similar in mutant and control littermates at different time points in culture development. (B) Representative phase plots (dV/dt vs V) from mature (> 21 DIV) mutant (orange) and control (black) neurons, exemplifying the effect of the $qv^{3J}$ mutation on the first, but not on the second phase of the AP. Arrows and letters indicate the quantifications of AP waveform used in C through F. (C-D) The quantifiers of the early AP phase, threshold and onset rapidness, increasingly diverged between $qv^{3J}$ and control during development. The difference between mutants and controls was significant already after the second week *in vitro*. (E-F) The second phase of the AP was unaffected by the mutation. The peak voltage and peak rate of rise have similar values in both mutants and controls. A and C –F Number of cells (animals, litters): $n_{mutant}$ = 36 (10,6), 44(9,6), 36 (6,4), 47 (7,5), 43 (6,4); $n_{control}$ = 70 (13,11), 96 (12,10), 78 (12,9), 41 (7,6), 50 (6,5). Displayed is the mean, shaded area represents the bootstrap 95% confidence interval.

**βIV-spectrin is lost early on, NaV channels cannot be stabilized or recruited**

In the AIS of control neurons we observed, at all studied time points, strong signals from antibodies against AnkG, βIV-spectrin and $Na_V$ (pan-$Na_V$). All antibodies used displayed minimal background staining, which allowed for a quantitative evaluation based on the immunocytochemistry. Indeed, in STORM imaging, the same staining protocol permitted single molecule detection (see Figure 6). Example images for the fourth week are shown in Figs. 3A and S3A. To quantify the spatial distribution of the immuno-signal, we obtained the fluorescence intensity profile along the axon (Figs. 3B, C; S3B, C), starting at the soma or, in the case of axons branching off a dendrite, starting at the last branching point. As expected from the electrophysiology results in controls, the $Na_V$ signal in the AIS increases over time, consistent with earlier reports (Yang et al., 2007). The fluorescence intensity of AnkG antibodies increased as well, similar to very recent observations (Yoshimura et al., 2017). When we turned to neurons from littermates homozygous for $qv^{3J}$, we first asked whether βIV-spectrin is still present at the AIS. Only at the earliest time point studied, at day in vitro 7, shortly after AIS assembly, we could detect a βIV-spectrin signal (Fig. S3D). Even then, the fluorescence intensity was much weaker compared to the control group (mean ± std of intensity along the first 30 μm of the AIS, 4.9 ± 1.2 [AU] in the mutants, n=4 cells; 105.6 ± 58.8 [AU] in controls, n=4 cells).

In contrast to the controls, $qv^{3J}$ neurons failed to accumulate AnkG or $Na_V$ in the AIS during development (Fig. 3B, orange profiles). Therefore the density of $Na_V$ at the AIS is lower in the mutant. A similar observation had been made for voltage gated potassium channels (Devaux, 2010). Given the low fluorescence signals in $qv^{3J}$, it was not always possible to unequivocally identify the axon as the one neurite that featured a stronger pan-$Na_V$ immunostaining in the proximal part. The intensity profiles in Fig. 3 and Fig. S3 represent the subset of $qv^{3J}$ neurons where AIS identification succeeded. When we combined pan-$Na_V$ and AnkG staining in a separate group of experiments, covering a shorter developmental period, we found that AnkG labelling allowed AIS identification even in cases where none of the neurites showed a distinctive pan-$Na_V$ signal. We attribute this reliability of AnkG staining to the fact that, unlike pan-$Na_V$, AnkG antibodies did not label soma and dendrites. This allowed us to identify the axon even in some of the cells, where the $Na_V$ label did not allow a clear AIS identification. The fraction of AnkG-positive but Na$_V$-negative AIS grew over time (Fig. 3D) from 15% at 11 days in vitro (DIV) to 66% at 19 DIV.

Remarkably, despite the strong reduction of Na$_V$ immuno-labelling at the AIS, the density in the soma appeared unaffected by the mutation (Fig. 3E). Therefore, electrophysiological assessment (Fig. 2A, B, F and G) and immunostaining both indicate that the structural and functional consequences of the qv$^{3J}$ mutation are restricted to the AIS, while the structural and electrical maturation of the somato-dendritic compartment is preserved.

In order to assess, whether our immunocytochemical findings allow a semi-quantitative characterization of Na$_V$ densities, we compared them with the electrical signature of Na$_V$ density. To this end, we recorded from 23 control neurons, which were then stained with pan-Na$_V$. When staining intensity in the AIS was plotted against the onset rapidness, a strong dependence was found (Fig.S3), similar to the relation between Na$_V$ density and onset rapidness in the model (Fig.1G, circles). Similarly, plotting staining intensity in soma against the peak of the second phase of the AP phase plot also revealed a high correlation (Fig.S3). This clearly indicates that the pan-Na$_V$ stain gave a quantitative estimate of the functional Na channels in the plasma membrane. Hence our imaging results (Fig. 3E) show that the Na$_V$ density in the somatic membrane was unchanged in the qv$^{3J}$ neurons. Furthermore, the axonal density of Na$_V$ in the mutant neurons at DIV 25-27 was about 75% reduced as compared to controls (Fig. 3C). This is only a lower bound of the actual reduction. As mentioned above, mutant neurons in which the Na$_V$ label was too weak to identify the axon were excluded from this analysis.

The observed early loss of βIV-spectrin suggests that the frame shift mutation towards the C-terminus removed an important interaction site that supports the retention of βIV-spectrin at the AIS. The subsequent loss of AnkG and Na$_V$ could be simply a consequence of the disappearance of βIV-spectrin, however, it is unclear how the remaining AnkG and Na$_V$ proteins are stabilized at the AIS, when βIV-spectrin is already lost. Furthermore, the complete loss of this key structural protein raises the question, whether the regular nanostructure, which is thought to rely on spectrin tetramers, is also destroyed in qv$^{3J}$.

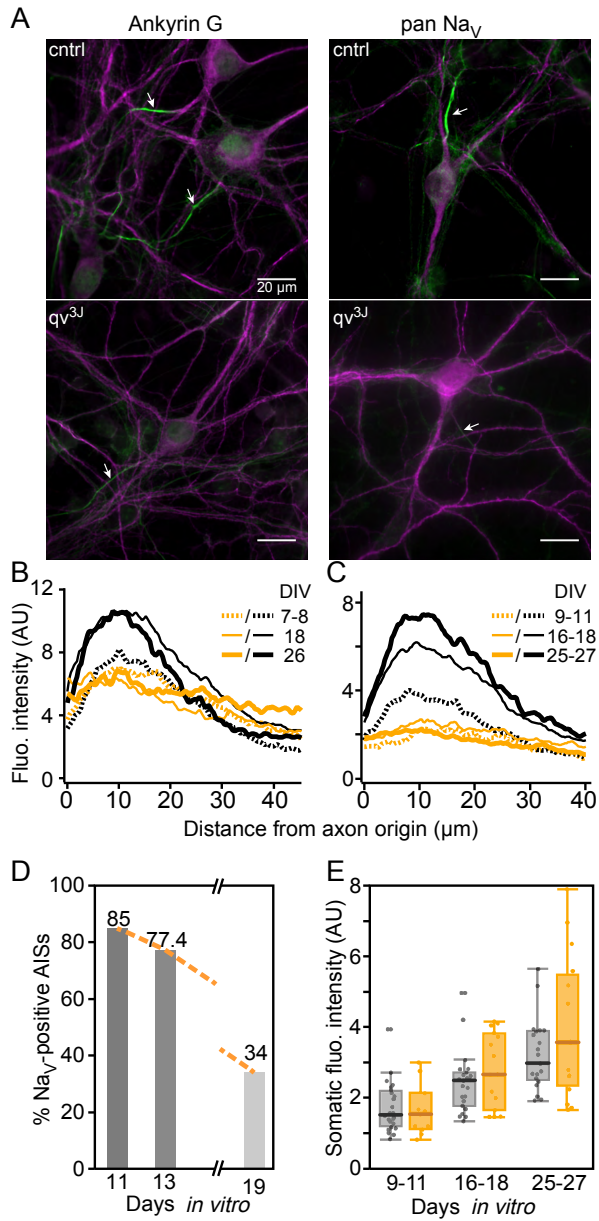

**Figure 3 Lower densities of Na$_V$ channels and AnkG in the AIS of mature qv$^{3J}$ mutant neurons.** (A) Hippocampal neurons, 26 days in culture, from mutant and control animals were labeled with antibodies against MAP2 (magenta) and either Na$_V$ channels (pan-Na$_V$) or AnkG (green). (B) Profiles of AnkG immuno-signal along the AIS, starting where it branches off the soma or a dendrite. Data from mutant are shown in orange, control in black at three developmental stages: 7-8 DIV, 18 DIV, 26 DIV. n$_{mutant}$ = 25 (2, 2), 17 (2, 2), 14 (2, 2); n$_{control}$ = 26 (3, 2), 32 (3, 2), 25 (4, 2). (C) As in B, but for pan-Na$_V$ immunolabelling: 9-11 DIV, 16-18 DIV, 25-27 DIV. n$_{mutant}$ = 13 (2, 2), 16 (2, 2), 19 (3, 3); n$_{control}$ = 38 (4, 2), 83 (11, 4), 32 (6, 3). In most mature mutant cells (>21 DIV) we could not identify an AIS based on its stronger pan-Na$_V$ staining. The line profile shown here was obtained from the subset of neurons that still demonstrated an evident staining. To reduce cluttering, only the averages are shown in B and C, confidence intervals are shown in figure S3. Note that AnkG and Na$_V$ channels fluorescence intensity at the AIS of mutant cells remains low throughout maturation (orange curves), but increase in control cells (gray curves). (D) In a separate data set, covering a smaller developmental period, cultures were double stained for AnkG and pan-Na$_V$. Here axons could be identified by AnkG label even when no neurite showed enhanced pan-Na$_V$ staining. Among the AnkG-positive axons, the fraction pan-Na$_V$ positive AIS' decreased during development; n = 13 (1, 1), 84 (4, 3), 35 (3, 2). (E) Average somatic pan-Na$_V$ fluorescence intensity is similar in control and qv$^{3J}$ mutant cells during development, suggesting it is unaffected by the mutation. n$_{mutant}$ = 11 (2, 2), 15 (2, 2), 16 (2, 2); n$_{control}$ = 24 (5, 2), 23 (10, 4), 21 (5, 2). AU, arbitrary units of camera signal, obtained under standardized conditions. Replication numbers refer to cells (animals, preps).

**Combined patch clamp and immunostaining confirms axonal initiation despite low channel density**

The results from electrophysiology and immunohistochemistry suggest that $Na_V$ densities are reduced only in the axon, not the soma. Biphasic AP shapes, indicative of axonal AP initiation, occur at those developmental stages (Fig. S2, DIV 25 and DIV 30) at which axonal $Na_V$ immuno-labelling becomes indistinguishable from somatic labelling (Fig. 3D). This already suggests that AP initiation can occur in axons with $Na_V$ densities far below control levels. For a direct comparison, however, electrophysiology and immuno-labelling need to be performed on the same preparations and cells. We thus set out to apply both techniques in the same sample to test most strictly, whether AP initiation can occur in axons with $Na_V$ densities far below control levels.

We recorded APs from 11 $qv^{3J}$ mutant neurons, aged 14 to 28 DIV and filled them with an Alexa dye through the patch-pipette (Methods). Immediately afterwards, the cultures were fixed and labeled with antibodies against AnkG and $Na_V$ channels. By this approach, we could identify the axon relying on the coincidence of AnkG staining and intracellular labelling via the patch pipette (Fig. 4 & S4, see supplemental video). In 4 out of the 11 neurons the AnkG were so weak that no identification was possible.

The staining and imaging protocols used here are identical to the ones used for figure 3. This allows for a direct comparison of the intensities. Indeed, the somatic pan-$Na_V$ signal is in the range of 2 to 4, as reported in figure 3E. The axonal pan-$Na_V$ signal of the filled neurons, however, is often lower than the typical axonal signal in the previous dataset (see fig. S3C for avg and 95% confidence intervals), due to the different identification criteria. The example in figure 4 B has an axonal pan-$Na_V$ signal between 0.5 and 1. This is less than half the average in figure 3C, but more importantly, this corresponds to an 85 to 95% reduction from the peak of the average intensity profile in control neurons (Fig. 3C). Despite the low axonal density, the APs in this neuron are clearly initiating away from the soma, evident from the biphasic phaseplot. One neuron with a similarly weak

pan-Na$_V$ signal lost the biphasic AP rise (Fig. 4C). This within-cell comparison of labelling and AP shape clearly confirmed that APs can still initiate at the axon even when Na$_V$ channel density is lower than the somatic density (Fig. 4A, B).

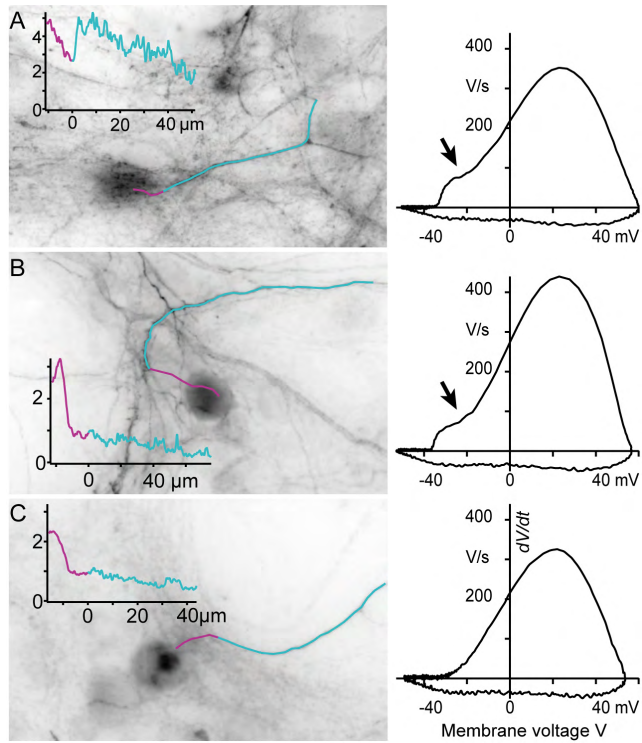

**Figure 4: Axonal initiation can occur even when channel density in the axon is lower than in the soma.** Immunolabeling and phase plot analysis in the same mature (24-28 DIV) qv$^{3J}$ neurons. (A-B) The biphasic phase plots demonstrate that APs were still initiated in the soma even when fluorescence intensity profiles indicate that the axonal Na$_V$ channel density was either slightly larger than in the soma (A: 1.5 times higher) or even lower than in the soma (B). (C) In this neuron, APs appeared to initiate in the soma as the phase plot shows no sign of any lateral current into the soma (compare figure 1B 99% reduction). The immunostaining in the same neuron reveals an axonal pan-Na$_V$ signal that is much lower than in the soma. The full data set comprising more neurons is shown in Fig. S4. Here, images were rotated for display purposes. A movie detailing the 3 fluorescence images used to identify the AIS and obtain the profile is in the supplemental data

**Information encoding in qv$^{3J}$ mutants**

So far, our experimental results on the AP initiation in qv$^{3J}$ mutants closely mirrored the behavior of the biophysical model under reduced axonal channel densities. We now set out to test the spike timing in qv$^{3J}$ mutants and controls. In analogy to the simulations, neurons were stimulated in whole-cell current clamp by injection of fluctuating currents with a correlation time of 35 ms, a value close to the typical membrane time constant (Fig. 2 A and B, Methods). The standard deviation was adjusted to obtain a firing rate of approximately 2 Hz (control ν=1.95 Hz (0.60) , n=11; qv$^{3J}$ ν=2.03 Hz (0.86), n=16). AP shapes and firing frequency showed no sign of adaptation throughout the 50s long stimulation episodes. Firing was highly irregular, the local variation of inter-spike intervals (Shinomoto et al., 2009) was on average lv=1.05 (SD 0.09) for controls and lv=1.06 (SD 0.08) for qv$^{3J}$ mutants. Fast spiking neurons, putative interneurons, had been excluded from analysis (Fig. S4).

To quantify the bandwidth of frequencies that the cells can encode in their AP firing patterns, we followed the approach of Higgs and Spain (Higgs and Spain, 2009) to calculate the dynamic gain. The overall shape of the dynamic gain was similar for both genotypes, the dynamic gain curves peaked around input frequencies of 30 Hz and fell for higher frequencies (Fig. 5C, D). However, in qv$^{3J}$ mutants this drop occurred at lower frequencies than it did in controls. This is evident from the non-overlapping bootstrap confidence intervals for the two populations (Fig.5D). To further characterize the difference, we compared the frequency at which the gain of individual cells drops to 60% of the peak value (Fig. 5C). This cut-off frequency was 36% smaller in qv$^{3J}$ mutants (Fig. 5E). These results are in line with the changes observed in models when reducing the axonal channel densities (Fig.1H, I).

A simpler assessment of AP timing precision is conditional firing rate, computed for two AP sequences fired in response to the same fluctuating input (Fig. 5F). As we used the same five stochastic stimuli for every neuron, we can characterize how precisely the population of neurons locks to the stimulus by computing the conditional firing rate for the two AP trains that a pair of neurons fired in response to the same input waveform. For threshold neuron models the conditional firing rate under frozen noise and the dynamic gain function are closely related (Tchumatchenko and Wolf, 2011). A tight locking between stimulus and AP initiation leads to little temporal jitter between the AP times of different neurons under the same stimulus (Fig. 5F) and hence to a high and narrow conditional firing rate (Fig. 5G). Averaging across all neuron pairs, we found that for the qv$^{3J}$ mutation, the average conditional firing rate has a smaller peak value and is broader: 9.4 ms full width at half maximum compared to 5.6 ms for the control neurons. Just as the reduced bandwidth of the dynamic gain, the broadening of the conditional firing rate reflects a less precise alignment between APs and the stimulus in the case of the qv$^{3J}$ mutant.

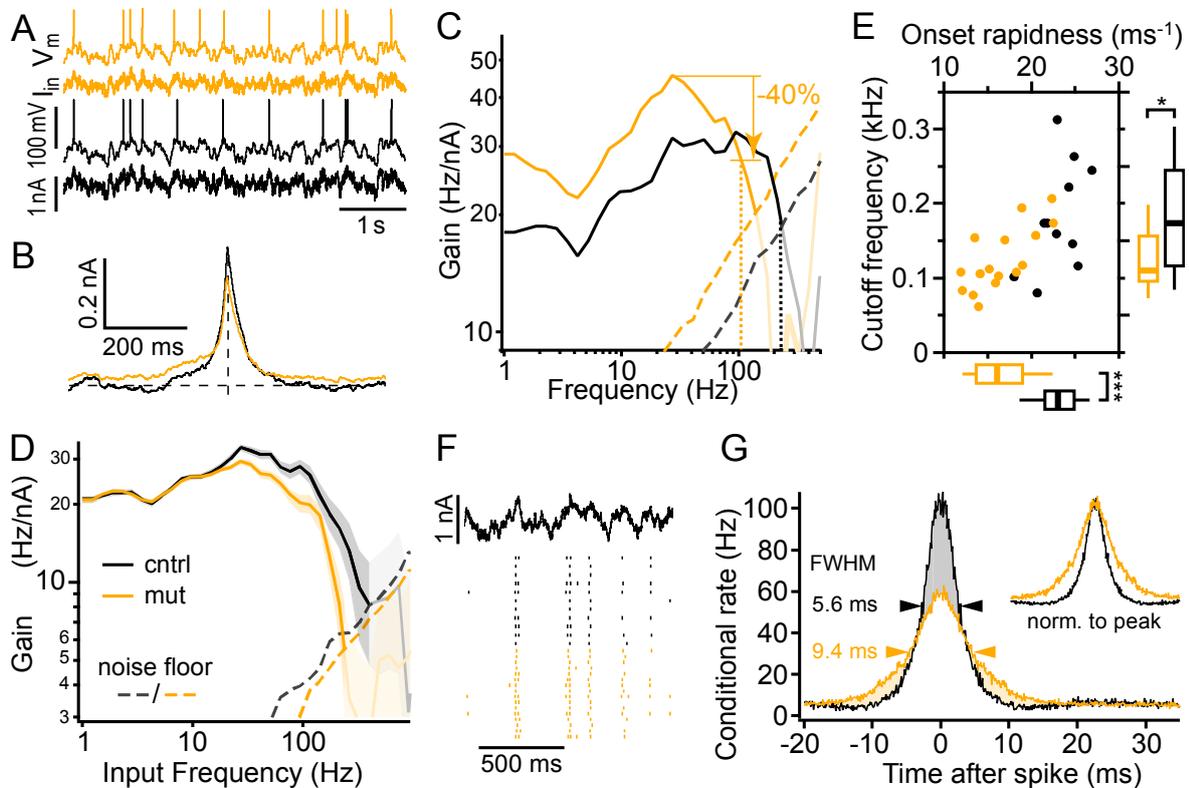

**Figure 5: Reduced precision of AP timing, bandwidth and slower AP onset in qv[3J] mutants.** (A) Fluctuating input current ($I_{in}$, see Methods) and the resulting membrane voltage ($V_m$) of one control neuron (black) and one qv[3J] mutant neuron (orange). (B) Spike triggered average current of the two neurons from A. (C) Dynamic gain for the two neurons (continuous lines) was considered significant up to the intersection with the noise floor (dashed lines). For calculations of gain and noise-floor see Methods. The cut-off frequency (dotted lines) for each gain curve was set as the point where the gain falls below 60% of its peak value. (D). Average dynamic gain of mature (>21 DIV) neurons from qv[3J] mutant mice and control littermates (wild type and heterozygous). Control: n=11 cells (2 mice from 2 litters); median age 31 DIV; 5,223 spikes; mutant: n=16 (4 mice from 4 litters); median age 29 DIV; 7,909 spikes). The frequency response function of mutant neurons (orange) drops at lower frequency compared with control (black) neurons (average with 95% confidence interval, see methods). Gain curves were considered significant until the intersection with the noise-floor (dashed). (E) Cut-off frequency plotted against AP onset rapidness (see figures 1 I and 2 D) for mutant and control cells. Box plots (median, quartiles and 10/90 percentiles) characterize the marginal distributions. The cut-off frequency was significantly lower in mutants (125.4 ± 10.6 Hz; mean ± SEM) than in controls (181.2 ± 21.8 Hz), *p = 0.0182, (2 sided student's t-test). The AP onset rapidness was significantly smaller in mutants (16.5 ± 0.9 ms$^{-1}$) than in controls (23.1 ± 0.8 ms$^{-1}$); ***p=0.000047 (Wilcoxon rank two tailed). (F) In response to frozen noise (top, stochastic stimulus 1, shown is the average fluctuation amplitude used across all 27 cells), 16 mutant and 11 control neurons fire APs locked to the stimulus. Shown is a 2 s long interval with a slightly higher than average activity. (G) Quantifying the precision of AP firing with the average conditional firing rate of pairs of neurons (see methods), the qv[3J] neurons show a broader peak, indicative of a reduced precision.

The experimental results on the functional consequences of reduced axonal channel densities closely match the behavior of the biophysical model (Fig.1), where all channel densities in the axon have been reduced, but no other changes occurred. Unlike the channels implemented in this model, ion channels in neurons operate in a context of other proteins and structural specializations. Most notably, $Na_V$ and their anchoring proteins in the AIS are arranged in a highly regular nano-structure (Xu et al., 2013; Zhong et al., 2014) of unknown functional relevance. To assess a potential relation of our functional findings to this nano-structure, we studied next, whether $qv^{3J}$ influenced the spatial order of axonal proteins.

**AnkG and sodium channels retain their periodic organization in $qv^{3J}$ AIS**

Although it is not clear, whether the regularity of the axonal cytoskeleton has any effect on electric signaling, one might argue, that the functional consequences of a reduced channel density could be offset by other functional modifications, possibly related to an altered ion channel clustering. Beyond the question of AP initiation, the maintenance of AIS cytoskeleton in the $qv^{3J}$ mutant is interesting in itself, because βIV-spectrin is thought to maintain the regular distance between actin rings (Fig. 1A).

We investigated the nanoscopic organization of AIS proteins using dSTORM (direct stochastic optical reconstruction microscopy), with Alexa 647 labelled secondary antibodies. As reported (van de Linde et al., 2011), the individual fluorophore localizations had a precision of 10-20 nm. In accordance with previous studies (Leterrier et al., 2015; Xu et al., 2013; Zhong et al., 2014) we found that AnkG and $Na_V$ are periodically spaced in the AIS, shown by a pronounced peak in the power spectrum of localization profiles at a spatial frequency of 1/190 nm (Fig. 6). Our data show that both AnkG and $Na_V$ retain their periodic distribution in the AIS despite the complete loss of βIV-spectrin in the mutant. The periodicity of AnkG seemed even slightly more pronounced compared to control cells, possibly due to the reduced protein density. At 19 DIV, when the majority of mutant AIS do not display a discernible pan-$Na_V$ labelling, the spectral power at 1/190 nm of the pan-$Na_V$ label is reduced in the mutant (Fig.6 B right panel). This break-down of regularity is likely due to the labelling sparsity. Lower spatial frequencies are over-represented as it would be expected when individual spots contain no pan-Na$_V$-label at all. To allow identification of the AIS despite the very weak Na$_V$ label, a double labelling with antibodies against AnkG was used, as for figure 3D.

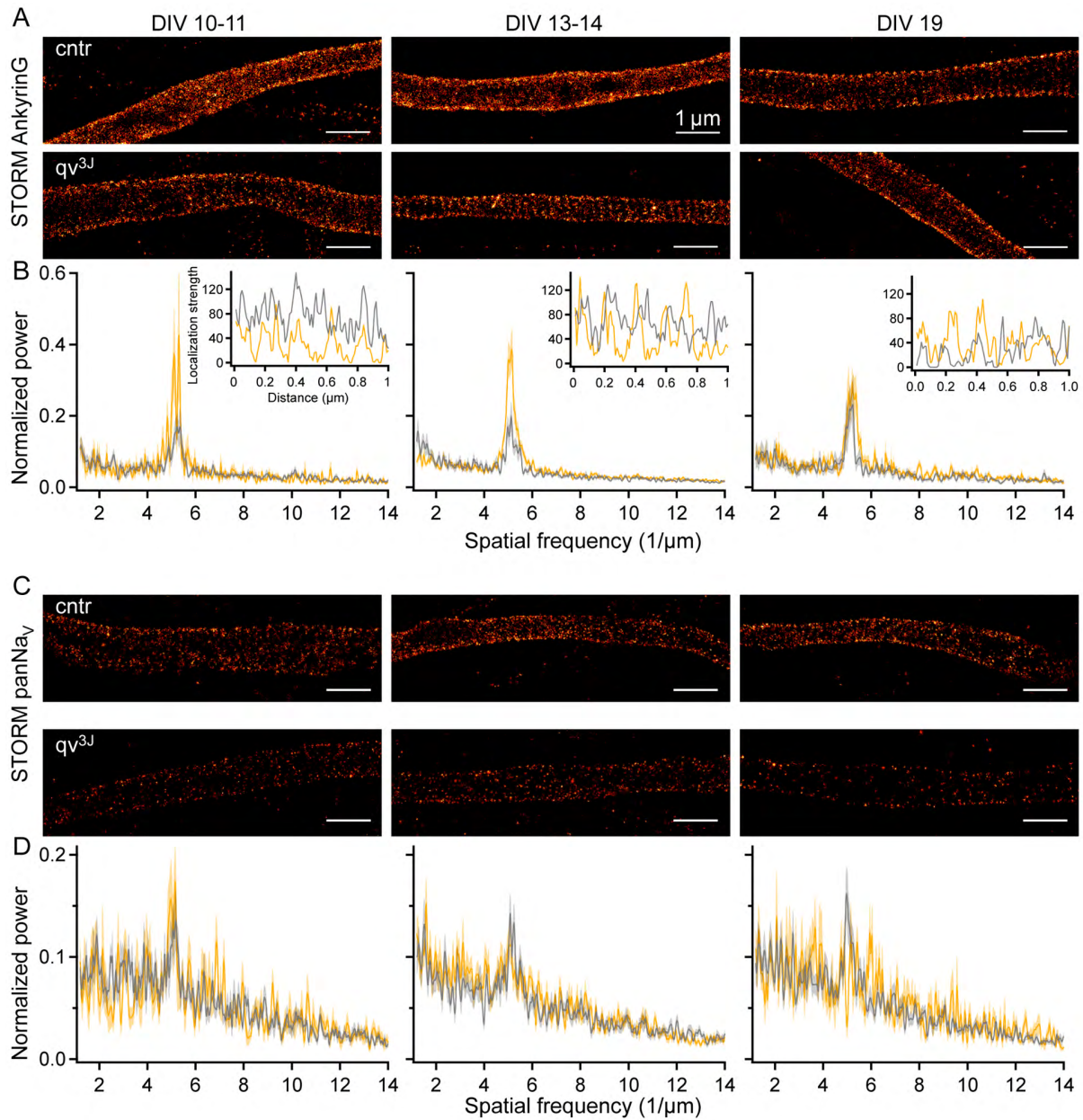

**Figure 6: The nanoscale spatial organization of AnkG and sodium channels appears unaffected by the $qv^{3J}$ mutation.** (A) dSTORM images of AISs of control (top) and mutant (bottom) neurons, labeled with antibodies against AnkG (N-terminus) at DIV 10-11, 13-14 and 19. Scale bar: 1µm. (B) Power spectra analysis of AnkG immuno-fluorescence profile along the AIS of control n= 17 (2, 2), 28 (3, 3), 14 (2, 2) and mutant n= 9 (1, 1), 53 (3, 3), 22 (1, 1), demonstrating periodic pattern in both populations at the 3 time points, with periodic length of about 190 nm. Examples of individual immunofluorescence profiles along 1µm segments are shown in the insets. (C, D) Same as A, B but with antibodies against $Na_V$, demonstrating periodic pattern with 190 nm periodicity in control population at all 3 time points. In $qv^{3J}$, periodicity at 190 nm is reduced and peaks appear at lower spatial frequencies, as less localizations occur and the grid is not fully decorated. $n_{control}$ = 20 (3, 2), 33 (3, 2), 33 (6, 2), $n_{mutant}$ = 10 (2, 2), 29 (4, 3), 12 (3, 2). Errors represent SEM. Replication numbers refer to cells (animals, preps).

**βII-spectrin preserves cytoskeleton nano-structure in the qv$^{3J}$ AIS,**

The presence and regular arrangement of AnkG and Na$_V$ channels, weeks after βIV-spectrin had dropped under the detection limit, was surprising to us. In light of the crucial role that βIV-spectrin plays for the localization of AnkG and Na$_V$ channels to the AIS (Komada and Soriano, 2002; Lacas-Gervais et al., 2004; Uemoto et al., 2007; Yang et al., 2004) one might have expected a disintegration of the regular arrangement. Based on the observations by Zhong and colleagues (2014), we speculated that a βII-spectrin could still be present in the mature AIS, thereby providing binding sites for AnkG. Indeed, antibodies against βII-spectrin labeled the AIS, as defined by the presence of AnkG, of neurons from mutant and control animals at all developmental stages tested. While the fluorescence intensity of βII-spectrin decreased by ~40% with development, it was still pronounced in mature neurons (> 20 DIV) (Fig. 7A, B). The intensity of βII-spectrin label in the AIS did not differ between qv$^{3J}$ mutants and control, indicating that βII-spectrin is not upregulated in the mutant to compensate for the complete loss of βIV-spectrin. In addition, βII-spectrin (labeled close to its C-terminal) demonstrated a highly regular spacing of 190 nm (Fig. 7C, D). Together these findings suggest that βII-spectrin constitutes part of the periodic actin-spectrin cytoskeleton of the mature AIS, to which AnkG and Na$_V$ channels bind. The regularity of the remaining cytoskeleton seemed unaffected by the loss of βIV-spectrin in the qv$^{3J}$ mutant. Apparently, the primary consequence is a reduction in the density of AnkG and the associated voltage gated channels, rather than a loss of cytoskeleton regularity.

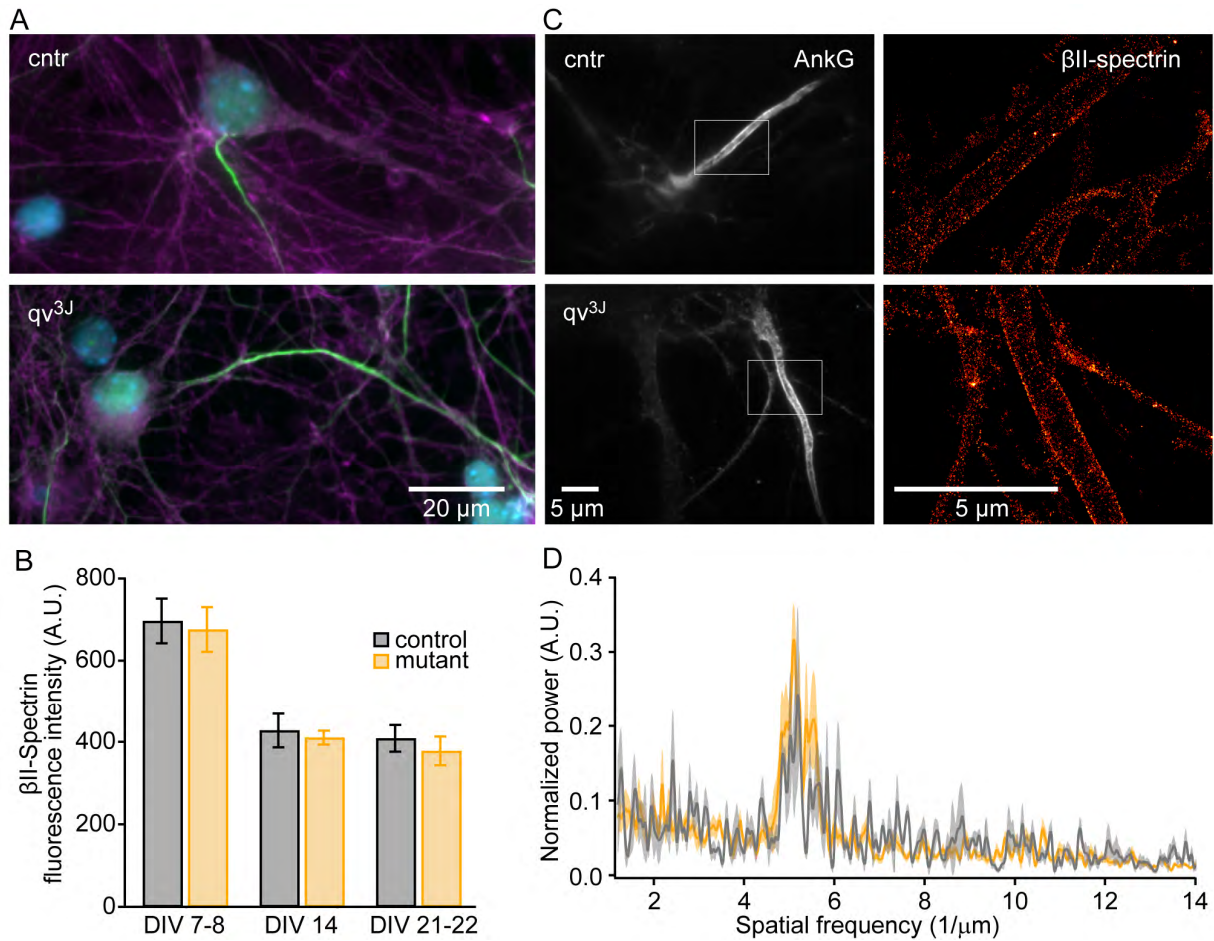

**Figure 7: βII-spectrin is expressed at similar levels in the AIS of mutant and control cells and is part of the periodic cytoskeleton**. (A) A culture of 8 DIV control (top) and qv[3J] mutant (bottom) neurons double labeled with antibodies against βII-spectrin (magenta) and AnkG (green) (C-terminal). Nuclei were stained by DAPI (blue). (B) βII-spectrin expression in control and qv[3J] mutant neurons at three maturation stages. The graph shows the mean fluorescence intensities in the AIS, within 50.0 µm from the soma or first branching point. Mutant and control cells showed similar βII-spectrin fluorescence intensities. βII-spectrin expression was reduced with development and shows no rescue for βIV-spectrin deficiency. $n_{control}$ = 36 (3, 2), 22 (2, 1), 27 (2, 2); $n_{mutant}$ = 32 (2, 1), 11 (1, 1), 22 (2, 2). Errors represent SEM. AU, arbitrary units. (C) Antibodies against AnkG (C-terminus) were used to identify the AIS (gray) in control (top) and mutant (bottom) cells (DIV 11 to 13) and a region of interest was chosen for βII-spectrin dSTORM imaging (red). (D) Power spectrum of βII-spectrin immunofluorescence profiles along the AIS of mutant n=14 (1, 1) and control cells n = 25 (3, 2) shows periodic organization in both populations, with periodic length of about 190 nm. βII-spectrin structural organization appears unaffected by the qv[3J] mutation. Replication numbers refer to cells (animals, preps). Errors represent SEM.

**Discussion**

The qv$^{3J}$ mutation in βIV-spectrin leads to a developmental loss of ion channel anchor proteins selectively in the AIS. By exploiting this experimental platform to disrupt the clustering of voltage dependent ion channels in the AIS, we essentially partially rewind the tape of AIS evolution. By utilizing a culture system, with its low fluorescence background, we are able to conclude that βIV-spectrin is absent from mutant AIS after DIV 10. Nevertheless, single molecule localization microscopy showed that AnkG and Na$_V$ antibodies are still arrayed with the 190 nm periodicity characteristic of the spectrin/actin membrane undercoat. The regularity is probably maintained by βII-spectrin (Zhong et al., 2014), which was present and not different in the AIS of mutants and controls. Levels of AnkG and Na$_V$ did drop substantially during the second and third week in culture, and ultimately neither could be labelled. The changes were isolated to the AIS since, over the same time span, Na$_V$ staining in the soma remained at control levels.

By determining the functional consequences of reduced axonal Na$_V$ and K$_V$ channel densities, we reveal the functional benefits afforded by the high AIS channel density. APs were initiated in the AIS, even when axonal Na$_V$ channel density was reduced to about 10% of control, to a level that was lower than in the somatic membrane. This experimental finding unambiguously demonstrates that AP initiation in the AIS does not require the high local channel density. However, the precision of AP timing was substantially compromised when axonal channel density was reduced. Taken together, our results indicate that while clustering of ion channels at the AIS is not the reason AP initiation is shifted from the soma into the axon, it is required for tight temporal locking between initiation of APs and transient increases in input, as evidenced by the increase in the bandwidth of information encoding. These experimental findings and conclusions are completely substantiated by computational results obtained using the most refined models of pyramidal neurons (Hallermann et al., 2012; Hay et al., 2013).

Was precise AP timing a functional advantage for the early vertebrates that first evolved ion channel clustering in the proximal axon? Fossil data and molecular phylogeny suggest that the last common ancestor of all vertebrates was similar to present day lamprey, and it is likely that it possessed camera-type eyes (Gabbott et al., 2016; Heimberg et al., 2010). Lampreys and mammals share basic neuronal infrastructure for complex behaviors, such as reward-based learning, visual processing for prey, conspecific and mate detection, and motor control (Grillner et al., 2018; Mikheev et al., 2006; Robertson et al., 2014). This suggests that the brains of stem vertebrates, whose neurons had an AIS, were complex and that they used multi-neuron populations to encode information. Our study reveals the unique features of the AIS which, by promoting precise AP locking to transient changes in a continuously fluctuating input, may have been a beneficial force in evolution of early vertebrates.

Our investigation is relevant to the many neurons in the mammalian central nervous system that operate in the fluctuation driven regime. AP firing in these neurons is generally asynchronous and seemingly stochastic. For large populations of neurons, information about their shared input is encoded in the population's collective firing rate. We assessed the precision of this encoding in the time domain by comparing the responses to repeated presentations of the same fluctuating input, and in the frequency domain by calculating the dynamic gain. Mutants performed much worse than controls in both assays. Recent theoretical studies showed that a larger dendritic compartment is associated with better encoding precision (Eyal et al., 2014; Ostojic et al., 2015) and that properties of the AIS need not be a factor. We find that the passive properties of mutant and control neurons are not different, which implies that if there are differences in their dendritic trees, these are minimal, and that this cannot account for the observed marked decreased AP timing precision in the mutant. Indeed, the strong reduction in the bandwidth of the dynamic gain seems to directly reflect the decrease in NaV density, as clusters of channels disassemble. Here, we directly demonstrate the dependence of dynamic gain on the physical properties of the axonal membrane at the site of AP initiation.

We are the first to show for on the level of individual neurons that AP onset rapidness is correlated with the bandwidth of information encoding. This dependence of encoding capacity on the active properties of the AP initiation is consistent with earlier theoretical studies (Fourcaud-Trocme et al., 2003; Huang et al., 2012; Naundorf et al., 2006, 2005; Wei, 2011). Ilin et al. (2013) used developmental changes in the AIS to probe the impact of its properties on encoding bandwidth, and reported that older neurons have wider bandwidth and a faster AP onset. However, this result could be interpreted as reflecting the growth of somato-dendritic size with age (Eyal et al., 2014). In our study, the specific molecular manipulation of AIS channel density provides definitive experimental evidence for the direct impact of channel density on the temporal precision of AP encoding.

It is likely that our findings in culture are valid for neurons in-vivo and in other experimental platforms. Firstly, the AIS develops normally in hippocampal cultures, which are widely used for molecular studies of the AIS (Galiano et al., 2012; Huang et al., 2017; Zhong et al., 2014). Secondly, key functional properties of our control neurons, AP waveform and encoding bandwidth, are almost identical to those reported for pyramidal neurons in neocortical layer 2/3 and layer 5 in acute slices (Boucsein et al., 2009; Ilin et al., 2013; Kondgen et al., 2008) and in-vivo (Doose et al., 2016; Tchumatchenko et al., 2011). Moreover, the biophysical model of AP initiation in a cortical pyramidal neuron (Hallermann et al., 2012) quantitatively reproduces our finding, suggesting that our findings are valid for neocortex.

We have shown that the functional benefit conferred by clustering of ion channels in the AIS is to promote precise encoding of information by populations of neurons. It is important to note that the properties of the AIS are not only altered by pathological conditions such as spectrinopathies (Wang et al., 2018); they may also be subject to modification as a component of neuronal plasticity under normal conditions (Grubb et al., 2011; Grubb and Burrone, 2010; Kuba et al., 2010; Lezmy et al., 2017). As yet, evidence for functional consequences of such plastic changes has been limited to measurement of changes in single cell excitability. In light of our findings, it will be interesting to measure the effect of such changes on the bandwidth of information encoding.

**Methods**

**Cell culture**

Hippocampi were isolated from newborn (P0) qv$^{3J}$ mice, collected in a serum-free Neurobasal–A medium (Life Technologies, 12349-015) with 100 mM HEPES buffer solution (Life Technologies, 15630-056) and digested with trypsin/EDTA 0.05%/0.02% (w/v, Biochrom) in PBS for 12 minutes at 37°C. The hippocampi were then transferred to Neurobasal-A medium, pipetted up and down for homogenization, and centrifuged for 2 minutes. Hippocampi from each newborn offspring were used to prepare a separate culture. For the dataset in Fig.S3E, hippocampi from E18 wildtype mice were pooled. For electrophysiology measurements and wide-field imaging, cells from the hippocampi of each newborn mouse were plated in one 8.8cm$^2$ cell culture dish (Nunclon, 153066), on 7 glass cover slips (Thermo Scientific, Menzel-Glaeser 10 mm #1), coated with 0.1 mg/ml poly-L-lysine (Sigma-Aldrich, P2636), in 2 ml Neurobasal-A medium supplemented with 1:50 B27 (Life Technologies, 17504-044), 1:400 glutamax (Life Technologies, 35050-038) and 0.01 µg/mL fibroblast growth factor (b-FGF, Life Technologies, 13256-029). For dSTORM imaging, the cells were seeded in 4-well tissue culture chambers on cover glass (Sarstedt 94.61990.402), one hippocampus per well, in supplemented medium. Cultures were kept in an incubator at 37°C in a humidified atmosphere of 95% air and 5% CO$_2$. One half of the culture medium was changed weekly with freshly prepared medium. Tail biopsies were used for genotyping and were afterwards related with the corresponding culture.

**Genotyping**

Mice were genotyped by PCR on tail biopsy samples using the following primers: forward, 5'-AGG CAG CGC CTT TGC TGC GTC-3'; reverse, 5'-TCC TGG TCA CAG AGG TCC TTA-3'. PCR mix contained 1.0 µl DNA, 0.2 µl of each primer, 0.4 µl DreamTaq DNA Polymerase (Thermo Scientific, EP0703), and PCR buffer (containing Tris-HCl (pH 8.8), ammonium sulphate (Sigma), MgCl$_2$ (Sigma- Aldrich), 2-mercaptoethanol (Merck), EDTA (pH 8.0) (Sigma), nucleoside triphosphates (dATP, dCTP, dGTP, dTTP) (Promega), BSA (Ambion - Life Technologies) and H$_2$O to final volume of 20 µl). PCR conditions were: 3 minutes of initial denaturation at 94°C, followed by 36 cycles of 30 seconds denaturation at 94°C, 30 seconds annealing at 60°C, and 60 seconds elongation at 72°C. Final elongation was performed for 7 minutes at 72°C. Enzymatic digestion was performed with StyI (10 U/µl) (New England Biolabs, R0500S). The PCR product was separated by 3% gel electrophoresis (+/+ 600 bp, +/- 600 bp + 350 bp + 250 bp, -/- 350 bp + 250 bp).

**Electrophysiology**

Whole cell recordings were performed in cells from cultures at different developmental stages *in vitro*. Extracellular medium contained (in mM) 134 NaCl, 4 KCl, 2 CaCl$_2$, 1 MgCl$_2$, 10 HEPES, and 20 glucose (pH 7.4 with NaOH, 290 mOsm). Synaptic blockers against AMPA, GABA-a and NMDA receptors (10 µM NBQX, 50 µM Picrotoxin and 100 µM APV) were added to the medium, and completely suppressed synaptic input and also abolished spontaneous activity in the culture. Patch pipettes were fabricated from PG10165-4 glass (World Precision Instruments) and contained K-gluconate based intracellular solution consisting of (in mM): 136 K- gluconate, 10 KCl, 5 NaCl, 0.1 EGTA, 1 MgATP, 0.3 NaGTP, 10 HEPES and 5 Phosphocreatine (pH 7.3 with KOH, 300 mOsm). Pipette resistance varied between 3 and 6 MOhm, yielding access resistances between 5 and 15 MOhm. In some experiments 100 µM (final concentration) Alexa 568 hydrazide (sodium salt, Life Technologies, A-10437) was added to the intracellular solution (Figs. 6 and S6). Whole cell patch clamp was performed at room temperature (26°C) using an EPC10 amplifier controlled by Patchmaster (both HEKA Elektronik). The recordings were sampled at 100 KHz (step pulses) or 50 KHz (fluctuating input). Appropriate bridge and electrode capacitance compensation was applied. Using Patcher's Power Tools (Dr. Francisco Mendez, Frank Würriehausen) the recordings were imported into Igor Pro (Wavemetrics) and analyzed with custom written routines. Measurements were corrected for calculated liquid junction potential (LJP) of 16 mV. All electrophysiological experiments were performed on cells from cultures derived from more than one embryo, from more than one litter. In figure S2 Data from a single litter are shown specifically to illustrate the clear difference between qv$^{3J}$ and control on the background of the substantial variability within a single litter. Neurons were included, when a complete dataset was available. For the data in figure 2, in rare instances a single measure was not available, for instance, when the estimation of cell capacitance was hampered by a particularly instable response to the hyperpolarizing test pulses. Provided all other measures, e.g. AP shape measures, were available, the neuron was included in those datasets in figure 2. This concerned at most 3 neurons (<5%) of any sub-group and in total only 2% of cells in figure 2, see also the uploaded data of figure 2.

**Current stimuli**

The subthreshold membrane properties and the characteristics of the action potential were obtained by injecting a series of 100 ms long current steps with increasing amplitudes. Between stimuli, the neurons were held for 5s at -76 mV (LJP corrected). For the data in Fig. 2, we analyzed the APs recorded for the smallest suprathreshold current for which an AP was generated within 10 ms after current onset.

Frequency response properties were investigated using input currents synthesized as Ornstein-Uhlenbeck (OU) noise with zero mean and correlation time of 35 ms. The stimulus was applied for 50 s in 5 trials, separated by 60-90 s intervals in which the cell was held at -76 mV. In each trial, the stimulus was generated with similar mean and standard deviation but different random seed (i.e. different initial values of the internal state of the random number generator). The same 5 realizations of the noise, created by the same set of random seeds, were used for the different cells. The standard deviation of the injected current was adjusted for each cell to achieve a firing rate of 2 - 3 Hz while the DC component was adjusted to warrant a membrane potential fluctuating around -60 mV. For most cells this DC component was no more than 20 pA.

**Dynamic gain calculation**

The frequency transfer function was calculated from responses to injected fluctuating current, using a method originally introduced by Bryant and Segundo (Bryant and Segundo, 1976) with modifications. AP time was registered when the somatic membrane voltage crossed 3mV, which corresponds to the steepest point on the AP waveform. The spike triggered average (STA) current was calculated for each cell from ~ 600 spikes by averaging stimulus waveform in a temporal window of 500 ms before and after the spike. To improve signal-to-noise ratio, the STA was filtered in the frequency domain using a Gaussian window w(f'), centered at frequency f' = f, with an SD of f/2π,

$$w(f') = \frac{1}{\sqrt{2\pi} \cdot (\frac{f}{2\pi})} \cdot exp\left[-\frac{1}{2}\left(\frac{f'-f}{f/2\pi}\right)^2\right]$$

.

This averages out neighboring frequency components of similar amplitude but random phase, i.e. noise. Deterministic frequency components with a phase that changes only mildly within the Gaussian window are not affected by this windowing. Thus,

$$STA_w(f) = \frac{\int STA(f') \cdot w(f') \cdot df'}{\int w(f') \cdot df'}.$$

If the train of APs is idealized as a discrete sequence of numbers with zero for empty samples and 1/dt for samples carrying an AP, then the product of the STA current and the firing rate ν equals the cross-correlation between input current and AP output. The frequency response function (or the dynamic gain), G(f), was then calculated as the ratio between the Fourier transform of this cross correlation $\widetilde{CC_{I\leftrightarrow AP}}$ and the Fourier transform of the auto-correlation of the input current $\widetilde{AC_I}$. The later is equal to the power spectral density (PSD) of the input current and for an OU process, the analytical expression of the PSD can replace the numerical auto-correlation.

$$G(f) = \frac{|STA_w(f)| \cdot \nu}{PSD(f)} = \frac{\widetilde{CC_{I\leftrightarrow AP}}}{\widetilde{AC_I}}$$

$$PSD(f) = \widetilde{AC_I} = \frac{4\,\tau_{corr}\,\sigma^2}{1 + (2\pi f \tau_{corr})^2}$$

Where $\sigma$ is the standard deviation of the input current and $\tau_{corr}$ is the correlation time of the noise.

To average the gain curves from N cells, we averaged the STA currents. To avoid over-representation of cells with a smaller input resistance, i.e. cells that require a larger amplitude of current fluctuations, we weighted the STA curves: $\overline{STA} = \frac{1}{N} \cdot \sum_{i=1}^{N} STA_i \cdot \frac{\bar{\sigma}}{\sigma_i^2}$ with the average input variance $\bar{\sigma} = \frac{1}{N} \cdot \sum_{i=1}^{N} \sigma_i^2$. The average cross-correlation was obtained by multiplication with the average firing rate:

$$\bar{\nu} = \frac{\sum_{i=1}^{N} n_i^{APs}}{\sum_{i=1}^{N} T_i^{rec.}}\,.$$

$$\overline{G(f)} = \frac{|\overline{STA}_w \cdot \bar{\nu}|}{\frac{4\tau_{corr}\bar{\sigma}}{1 + (2\pi f \tau_{corr})^2}}$$

For each neuron and for the population average we calculated the confidence intervals of the gain curve as well as the noise floor by balanced bootstrap. The confidence interval at a given frequency $f'$ was defined by the 5th and the 95th percentile of $G_{BST}(f')$ for 200 bootstrap gain curves calculated from 200 random samples of actual AP times. The noise floor at a certain frequency is understood as 95th percentile of $G_{BST}^{rnd}(f')$ calculated not from measured but from random AP times. To obtain random AP times without changing the statistics of the AP time series, we applied a cyclic shift of the injected current by a random value larger than 5 correlation times. This results in a random triggered average of the input which replaces the STA current in the calculations for $G_{BST}^{rnd}$. Neurons were excluded from the analysis, if signs of non-stationarity in AP shape or other signs of cell instability, such as a drifting baseline potential, were detected.

**Conditional firing rate**

Each neuron in the dataset of dynamic gain and conditional firing rate analysis was stimulated with the same five stochastically changing current stimuli $I^1$ to $I^5$ (see above: current stimuli), only the fluctuation amplitude was adjusted between neurons to achieve firing rates of approximately 2 Hz. A given neuron m fires the AP sequences $s_m^1$ to $s_m^5$ in response to those stimuli. The conditional firing rate between neurons m and k is defined as the average over the conditional firing rates of the five trials:

$$\nu_{cond\ m,k}(\tau) = (\nu_m \cdot \nu_k)^{-1/2} \sum_{i=1}^{5} \langle s_m^i(t) \cdot s_k^i(t+\tau) \rangle$$

To obtain the average conditional firing rate of the neurons from control of mutant mice, we averaged

$$\nu_{cond}(\tau) = \frac{2}{N_{cells} \cdot (N_{cells} - 1)} \sum_{m=2}^{N_{cells}} \sum_{k=1}^{m-1} \nu_{cond\ m,k}(\tau)$$

**Action potential characterization**

The somatic AP waveform was characterized with the following set of parameters: the threshold voltage ($V_{thresh}$) was estimated as the value of the membrane potential when dV/dt crosses 25 V/s; the AP onset rapidness was calculated as the slope of the phase plot at the threshold voltage. The peak potential was the maximal value of V and the peak rate of rise was the peak value of dV/dt. This rate of voltage change is proportional to the sum of all currents that charge the capacitance C, i.e. $\frac{dV}{dt} = \frac{I}{C}$. At the peak of the second phase of AP depolarization, the main current source charging the soma are somatic sodium channels. Therefore, the peak rate of rise is indicative of the somatic sodium channel density. The relation is not exact, as lateral currents into the dendrites reduce somatic dV/dt.

Great care was taken to properly compensate the fast capacitance since it has an influence on the shape of the phase plot.

**Simulations**

A multi-compartment model, developed by Hallermann and colleagues (2012) to capture the properties of AP generation in pyramidal neurons in brain slices, was used for the simulations of AP initiation and spike time precision (https://senselab.med.yale.edu/modeldb/ ShowModel.cshtml?model=144526). The properties of all active conductances and their spatial distribution were left unchanged from the original model. To obtain the transfer functions for different axonal channel densities, many millions of seconds had to be simulated. To reduce the computational load, the model morphology was compacted: the basal dendrites and the apical dendritic branch and the initial axon were each represented by a single process with adjusted geometry, two axon collaterals are branching off the main axon. Compacting the morphology did not alter the characteristics of the model for AP waveform or transfer function, but greatly expedited simulations. For some simulations, we exchanged the somatic sodium channel model form Schmidt-Hieber and Bischofberger against a similar model we had derived from our own measurements, the AP waveforms and transfer functions changed only marginally, this version of the model was used for the results presented here.

To obtain the transfer function, the model is driven by somatic injection of a fluctuating current, derived from an Ornstein-Uhlenbeck process with correlation time 35 ms. The mean and standard deviation of the process are chosen to obtain a firing rate of 2 Hz and a coefficient of variation of the inter-spike interval around 0.85. This reflects a fluctuation-driven state and closely matches the experimentally obtained firing statistics. Each simulation is 200 s long with sample interval of 0.025 ms, 250 such 200 s simulation are combined to obtain approximately $10^6$ spikes for each set of axonal channel densities. AP time is set when the somatic voltage crosses +8 mV. From AP times and the input current we calculate the spike triggered average input (STA input), which represents the cross-correlation of input and AP output. The transfer function TF is calculated as the ratio of the Fourier transformations of the cross-correlation (STA input) multiplied with the firing rate $f$ and the auto-correlation of the input. The latter corresponds to the power spectral density (PSD) of the input:

$$TF = \frac{FFT(STA) \cdot \nu}{PSD_{OU}}$$

**Immunocytochemistry**

Neurons were washed twice in PBS and fixed with 4% formaldehyde in PBS for 8 minutes at 4°C and then washed in PBS. Permeabilization was performed by 5-minute incubation with 0.5% Triton X-100 in PBS, followed by 5-minute incubation with 0.1% Tween in PBS. The cells were then incubated for 1.5 hours in 3% BSA and 0.1% Tween in PBS (blocking solution) and then in primary antibody diluted in blocking solution at 4 °C, overnight. After washing, secondary antibodies were diluted in blocking solution and applied for 45 minutes. For wide-field imaging, the coverslips were mounted on microscope slides using Prolong Gold antifade with or without DAPI (Life Technologies, P36935 or P36930). For dSTORM imaging, the cells were post fixated with 4% formaldehyde in PBS for 5 minutes, washed 3 times with PBS and stored with PBS at 4 °C.

The primary antibodies that were employed in this study were: mouse monoclonal anti-sodium channel (Pan-Na$_V$) 1:900 (Clone K58/35; Sigma-Aldrich, S8809-1MG), goat polyclonal anti-βIV spectrin antibody 1:200 (ORIGene, TA317365, targeting N-terminal sequence aa2-14), rabbit polyclonal anti-βIV spectrin antibody 1:200 (ATLAS, HPA043370, targeting a centrally located sequence), rabbit polyclonal anti-AnkG antibody 1:200 (Santa Cruz Biotechnology, sc-28561, targeting C-terminal sequence), mouse monoclonal anti-AnkG antibody 1:200 (Santa Cruz Biotechnology, sc-12719, targeting N-terminal sequence), chicken polyclonal anti-Map2 1:2000 (Abcam, ab5392), mouse monoclonal anti-βII spectrin 1:200 (Santa Cruz Biotechnology, sc-136074) .

The following secondary antibodies were used (1:1000). For wide-field imaging: Alexa Fluor 488 goat anti-mouse (Life Technologies, A11029), Alexa Fluor 647 donkey anti-goat (Life Technologies, A21447, Alexa Fluor 488 goat anti-rabbit (Jackson ImmunoResearch 111-545-003), Alexa Fluor 647 goat anti-rabbit (Life Technologies, A21245), Alexa Fluor 647 goat anti-chicken (Life Technologies, A214469), Alexa 488 Donkey anti-chicken (Jackson ImmunoResearch 703-545-155). For dSTORM imaging: Alexa Fluor 647 goat anti-mouse (Life Technologies, A21236).

**Wide-Field imaging**

Images were recorded using an inverted widefield microscope (Olympus IX-71) equipped with a water-immersion objective lens (UPlanSApo, 60x magnification, NA 1.2, Olympus). Violet (emission maximum at 390 nm), cyan (emission maximum at 475 nm), yellow (emission maximum at 603 nm) and red (emission maximum at 634 nm) LEDs (Spectra X light engine, Lumencor) were used for excitation of DAPI, Alexa 488, Alexa 568 and Alexa 647, respectively. The fluorescence light was separated from the LED light with a dichroic beam splitter (59004bs, Chroma Technologies) and additional excitation filters (BLP01-405R, Semrock for DAPI; FF01-520/35, Semrock for Alexa 488; and BLP01-594R Semrock for Alexa 568, BLP01-635R Semrock for Alexa 647) before being imaged on an electron multiplying CCD camera (EMCCD; DU-897-CS0-BV, Andor) with an effective pixel size of 160 nm; temperature of detector -50 °C.

Images were analyzed using the image line profile operation (with line width of 5 pixels) Igor Pro (Wavemetrics). Units refer to CCD pixel readout under strictly constant settings and parameters for immunostaining and imaging protocols within a dataset. "Arbitrary" units refers solely to the fact that the CCD's readout magnitude is arbitrary *overall*, but not within or between measurements. Specifically it is not possible to compare fluorescence intensity units between different antibodies or detection protocols (STORM and widefield), but as the confidence intervals on repeated measurements with identical antibody batches and identical staining and imaging protocols show, there is good comparability within those measurements (Fig S3 B,C). Part of the imaging process is selection of suitable samples. We only used images from cells, where the entirety of the quantified axonal section is in the same focal plane. If the axon grows on top of other structures the cell was rejected. Replication numbers for individual cells, cultures derived from individual embryos and the number of litters are given in the figure legends where applicable.

**dSTORM imaging**

The imaging buffer used was 10 mM TRIS containing 100 mM cysteamine hydrochloride (Sigma-Aldrich, M6500), 4.0 mg/mL glucose oxidase (Sigma-Aldrich, G0543), 0.57 mg/mL catalase (Sigma-Aldrich, C40-100MG), and 10 % glucose (Sigma-Aldrich, 49158-1KG-F). pH was adjusted with 0.5 M NaOH to pH 8.3-8.5. Chamber wells were completely filled with buffer and sealed bubble-free (air-free) by a regular coverslip. Images were recorded using an inverted TIRF microscope (Olympus IX-71) equipped with an oil-immersion objective lens (Olympus, UApoN, 100x magnification, NA 1.49, TIRF) and a 647 nm laser (PhoxX 647, 140 mW, Omicron Laserage, Germany). A quad-edge dichroic beam splitter (Di01-R405/488/561/635, Semrock) and a quad-band excitation filter (FF01-446/523/600/ 677, Semrock) were used to remove the laser light before imaging the fluorescence on an electron multiplying CCD camera cooled to -50°C (EMCCD; DU-885-CS0-#VP, Andor) with an effective pixel size of 80 nm. Raw movies typically contained 3000 - 4000 images recorded at 30 - 90Hz. Images were analyzed using rapidSTORM and custom written routines in Matlab.

**Acknowledgements:** This study was supported by the BMBF (01GQ0922), GIF (906-17.1/2006), VW foundation (ZN2632), BCCN (01GQ1005B), BFNT (01GQ0811) and the Max Planck Society

**Author Contributions:** Conceptualization: M.J.G., F.W.,A.N.,E.L.; Data Curation: E.L., M.D.; Formal Analysis: E.L., A.N., M.D., BF; Funding Acquisition: M.J.G., F.W., A.N., JE; Investigation: E.L., M.D., BF; Methodology: A.N., BF, JE, F.W., M.J.G.; Software: A.N., BF; Supervision: A.N., JE; Validation: A.N., E.L.; Visualization: E.L., A.N., M.D.; Writing: E.L., A.N., M.J.G., F.W.

**Declaration of Interests:** The authors declare no competing interests

## Supplementary Figures

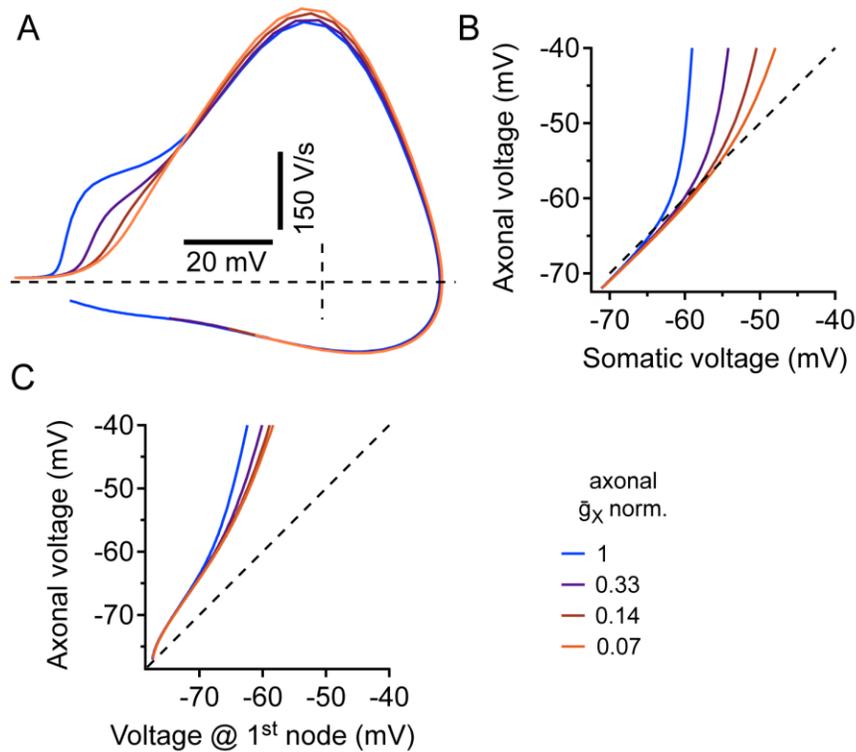

**Figure S1 Axonal AP initiation without channel density gradient in a model from Hallermann, Kock, Stuart and Kole (Hallermann et al. 2012)** (A) Phaseplots of somatic AP waveforms obtained for different levels of axonal channel densities. The density of $Na_V$ and $K_V$ channels has been reduced to the indicated level, compared to the original model parameters. Please note that the lowest density corresponds to $1/14^{th}$ of default at which somatic and axonal $Na_V$ channel densities are identical. (B-C) The comparison of isochronal values for voltages at the soma, AIS and $1^{st}$ node show that at all tested channel densities, APs initiate in the AIS.

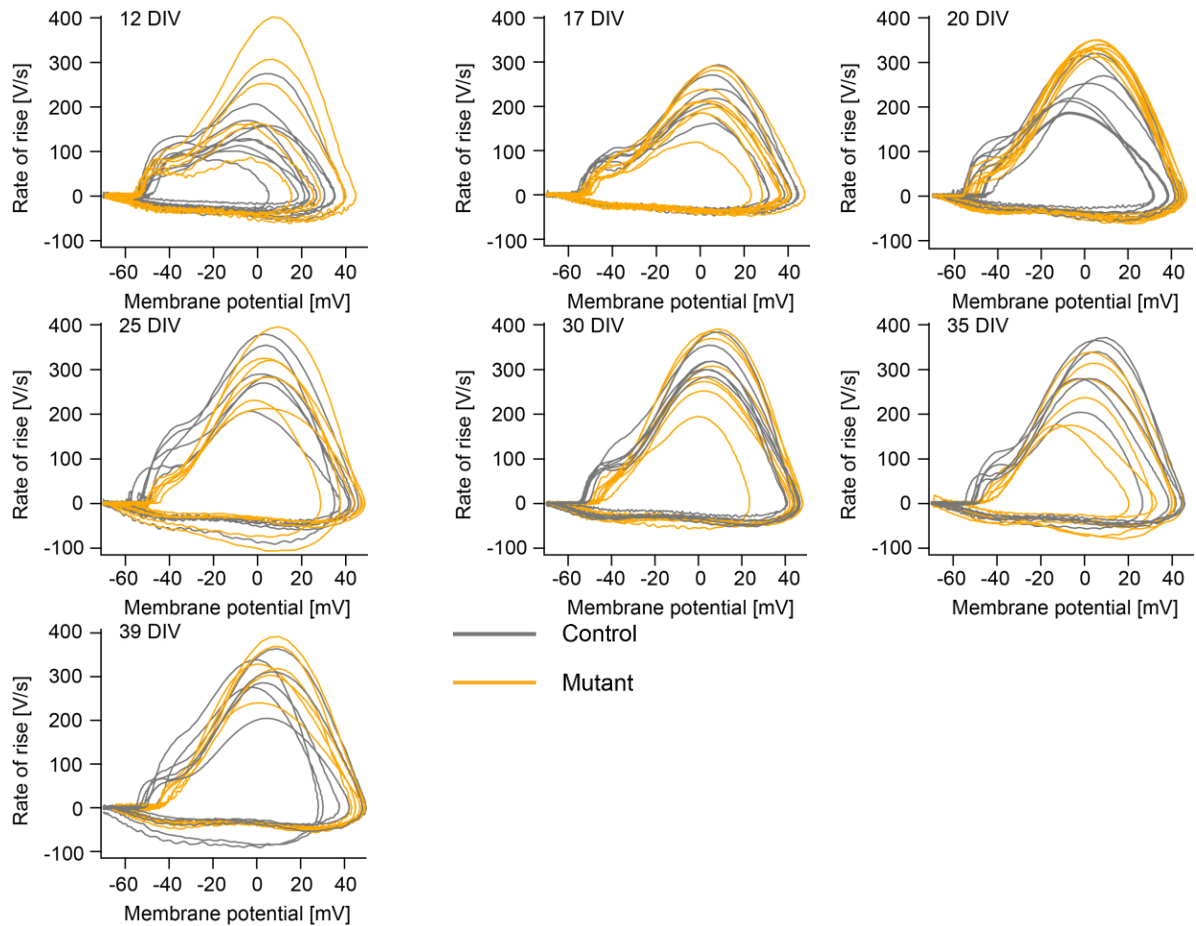

**Figure S2:** APs from qv[3J] mutant neurons lose their rapid onset with maturation. Phase plane analysis of APs recorded from mutant (orange) and control (gray) cells between day in vitro (DIV) 12 and 39 shows that at early developmental stage the AP waveform of mutant cells was biphasic as in control cells. Later during development the AP onset in mutant cells became slower and the biphasic shape was lost in an increasingly large fraction of cells. The cultures used for this data series are from two animals (one control, one mutant of a single preparation (i.e. one litter), with the exception of DIV12, where some recordings of a second control animal of that litter are added to display the full range of variable phaseplots at that stage. In each plot the APs were recorded on the same day under identical conditions from the different cultures. The cell population included regular- as well as fast-spiking neurons, which can be distinguished by their higher, i.e. more negative, repolarization rate.

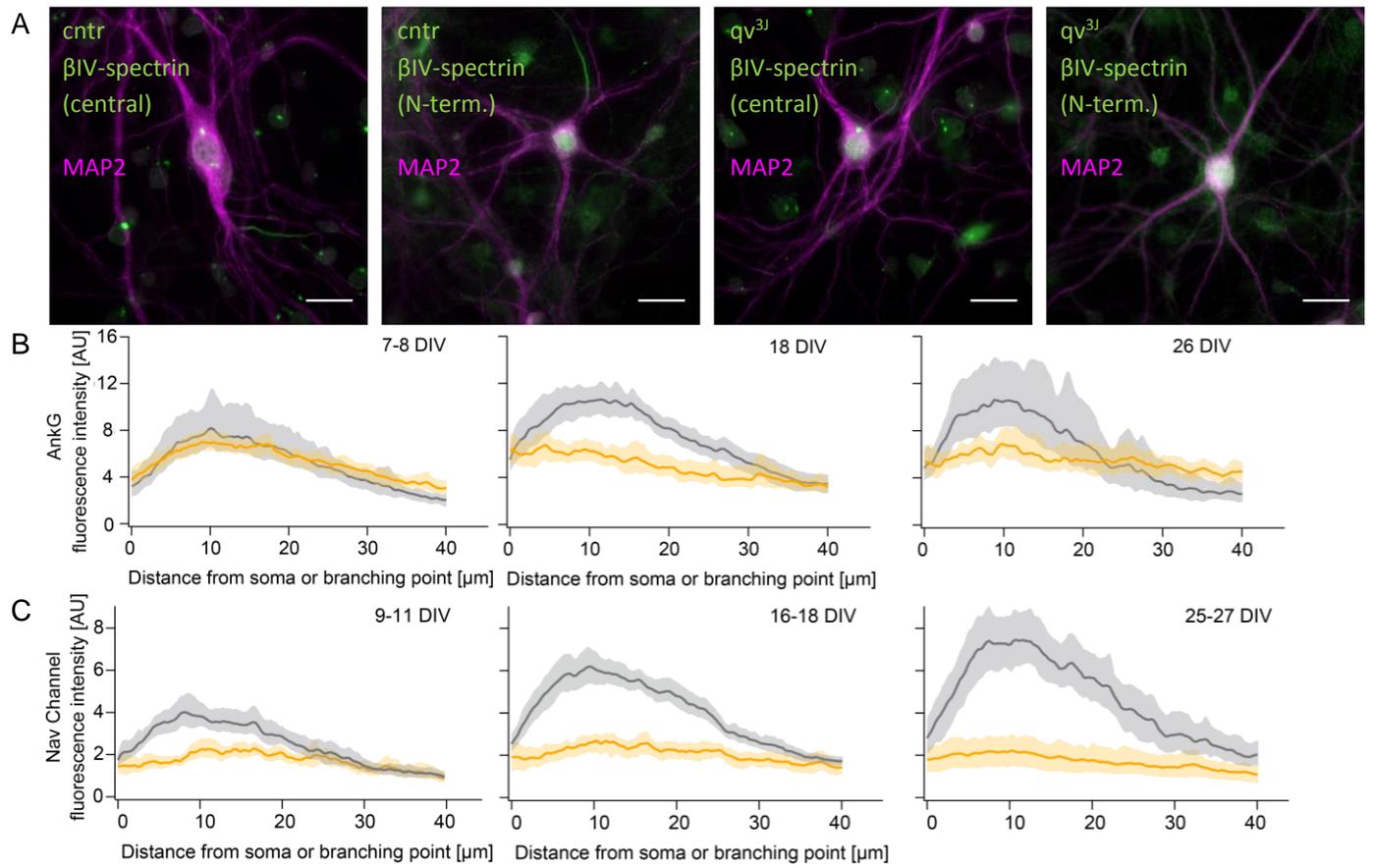

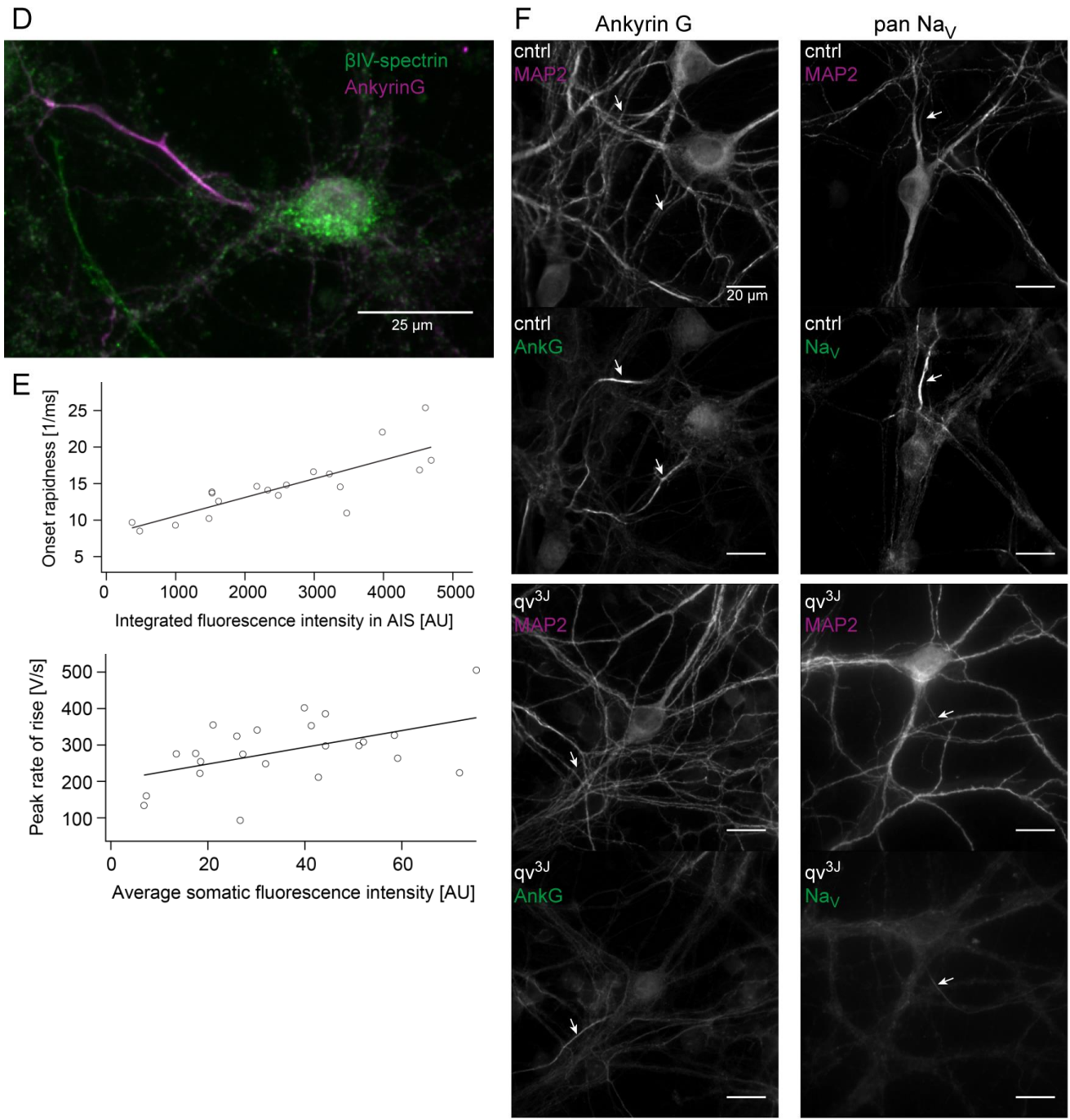

**Figure S3 βIV-spectrin is absent in qv³ᴶ mutants after first week of development, AnkG and pan-Na_V are significantly reduced** (A) Immuno-labelling with two antibodies against non-mutated regions of βIV-spectrin (see methods) show reliable AIS labelling in all control but not qv³ᴶ neurons. Scale bars are 20 µm. The example images here are from DIV 26. (B-C) Detailed presentation of the immuno-fluorescence profiles for AnkG and pan-Na_V from figure 3B & C). Here the average and the bootstrap 95% confidence intervals are shown for the fluorescence along the AIS, starting at the first branching point or the soma. (D) Example of a qv³ᴶ neuron at DIV 7 stained for AnkG (C-terminal antibody) and βIV-spectrin (N-Terminal, OriGENE Antibody), demonstrating low levels of βIV-spectrin are still present in the AIS at this developmental stage, indicated by the weak white signal in the AIS. (E) 23

neurons from wildtype mice were patched to measure the AP waveform and subsequently fixed and stained with pan-Na$_V$ antibodies. The onset rapidness is strongly correlated (n=19, r=0.81, p<0.025) with pan-Na$_V$ intensity in the AIS (10-30 µm). Four cells with multiple AIS have been excluded from this analysis. Averaged somatic label intensity is correlated with the peak rate of rise (n=23, r=0.49, p<0.025). (F) The individual color channels of the composite images from figure 3A are displayed separately.

| Alexa 568 filling | pan-Na$_V$ | AnkG | Phase plot |
|---|---|---|---|
| 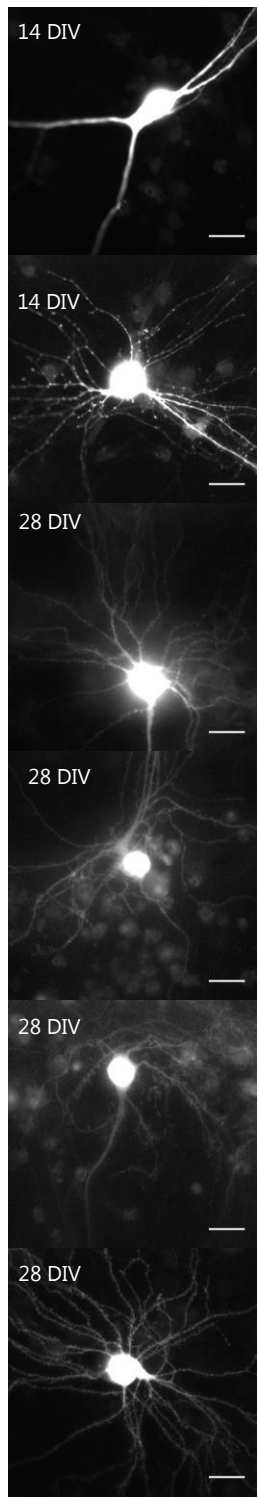 | 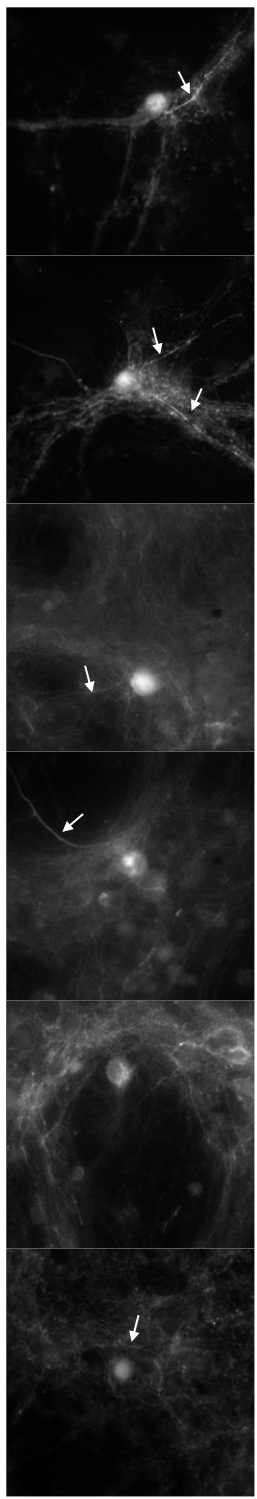 | 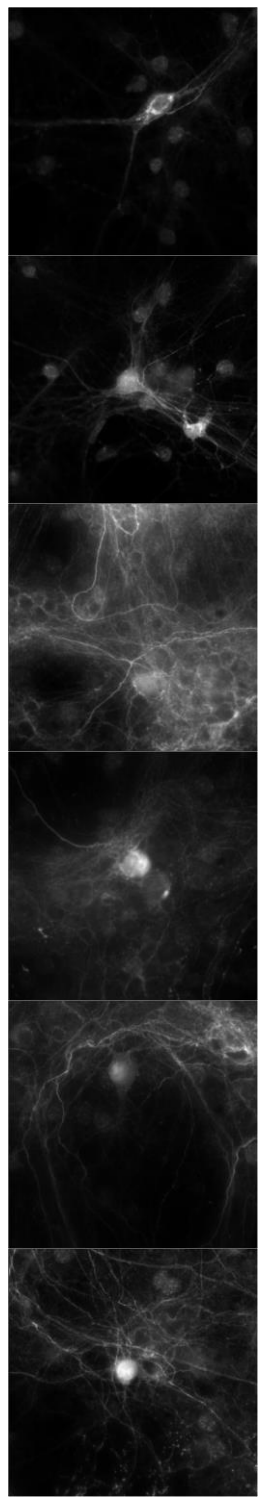 | 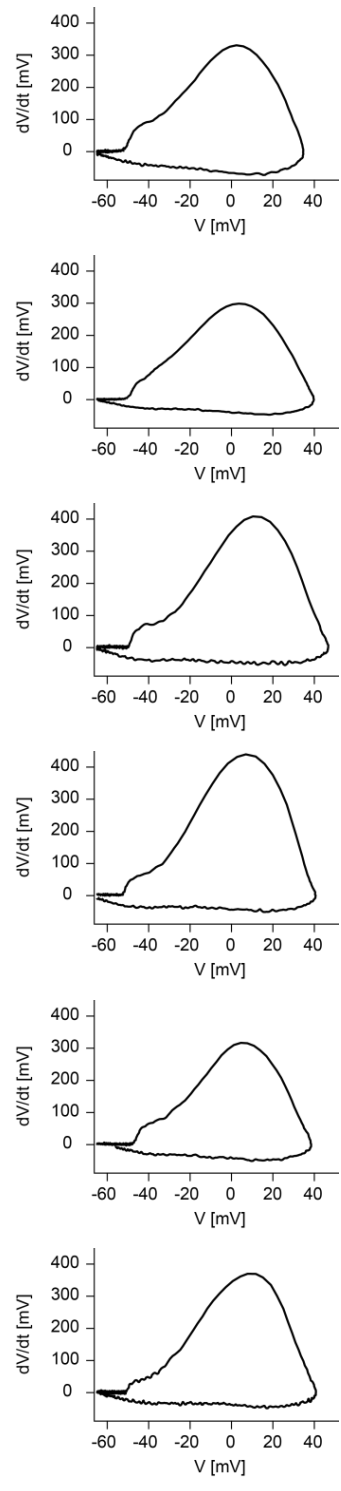 |

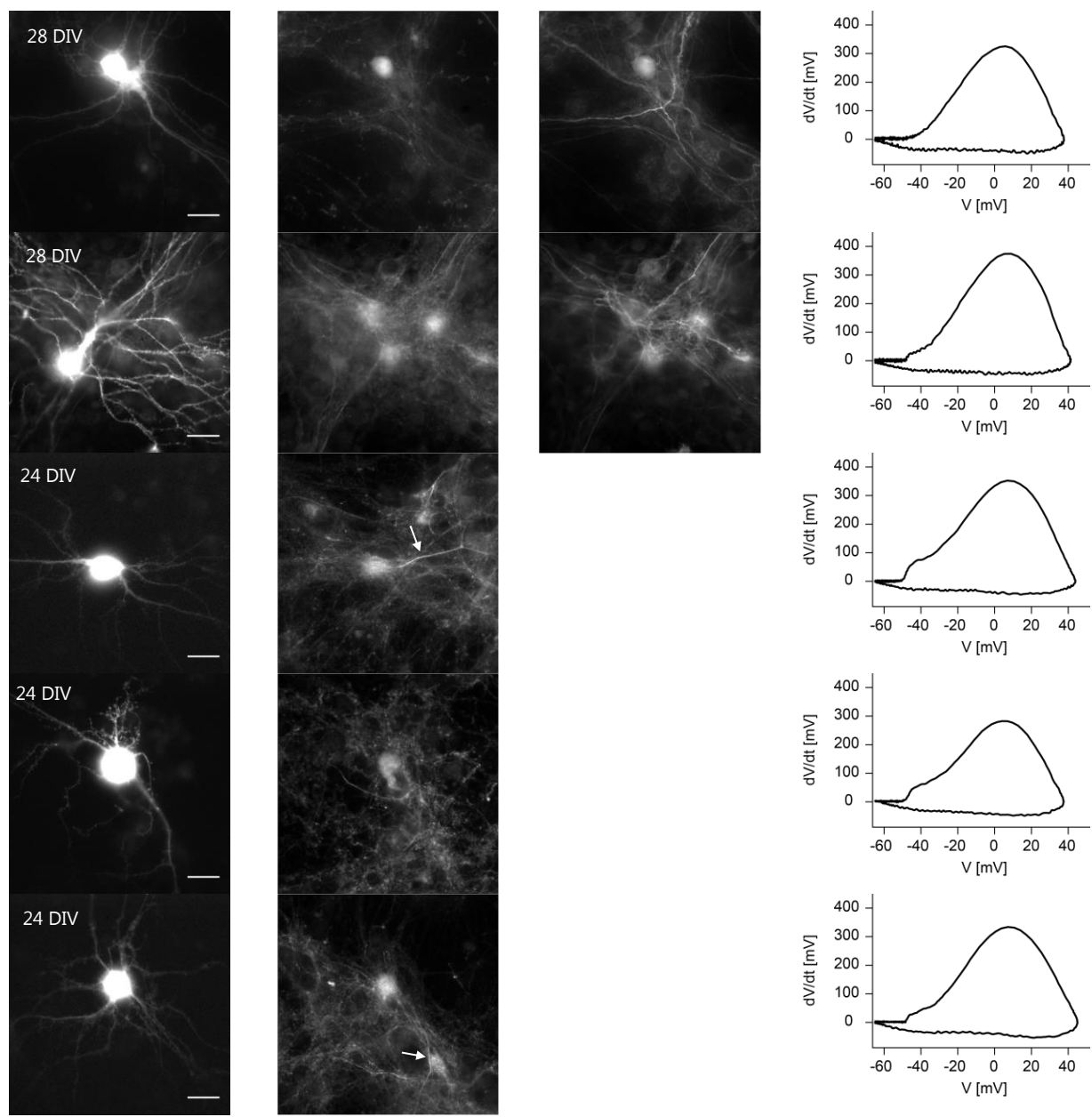

**Figure S4: Immunolabeling and phase plane analysis of homozygous qv$^{3J}$ mutant neurons at ages 14 - 28 DIV.** The 11 cells (from 2 animals from 2 litters) were filled with Alexa dye (left column) during whole-cell patch clamp recordings and were afterwards labeled with antibodies against pan-Nav channels and AnkG (excluding 3 cells) (second and third columns). White arrows mark the AIS in cases where we could identify it based on matching Alexa staining and immunolabeling with AnkG (see also the supplementary animation). In some cases we were not able to determine its location. Phase plots of APs recorded from the cells are shown on the right column. Scale bar: 20 µm. Note that each image of AnkG and pan-Nav labeling uses a different intensity scale (auto-scaled) and therefore their fluorescence intensities are not comparable.

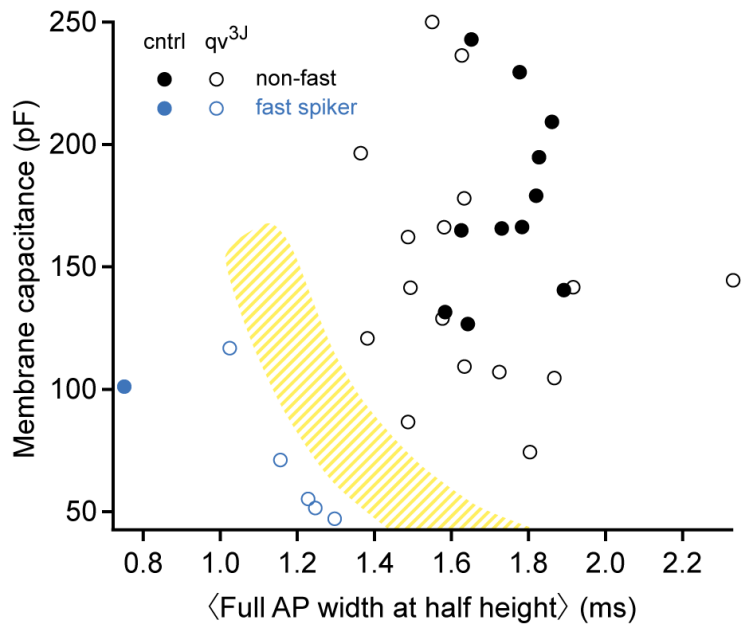

**Figure S5: Identification** of **putative fast spiking neurons** In cultures from control mice, fast spiking neurons were reliably identified by very brief APs, due to rapid repolarization. Examples can be seen in two phase plots in figure S2, DIV 39. In addition, fast spiking neurons tended to be relatively small, reflected in a small membrane capacitance. The population of neurons from mutant mice was slightly less homogeneous in their properties, therefore the fast spiking neurons were identified using a plot of membrane capacitance versus AP duration, measured as full width at half maximal depolarization. The hatched area in this plot separates neurons that are regarded as fast spiking neurons (lower left, 5 mutant 1 control) from neurons that are regarded as regular spiking neurons. This identification is supported by the neurons' firing patterns in response to constant current injections (not shown).